\documentclass[]{spie}  

\usepackage{amsmath,amsfonts,amssymb}
\usepackage{graphicx}
\usepackage[colorlinks=true, allcolors=blue]{hyperref}
\usepackage{makecell}
\usepackage{colortbl}
\usepackage{array}
\usepackage{afterpage}
\usepackage{multirow}

\usepackage[table]{xcolor}

\def\arcdeg{\mbox{$^\circ$}}%
\def\arcmin{\mbox{$^\prime$}}%
\def\arcsec{\mbox{$^{\prime\prime}$}}%

\definecolor{eric}{rgb}{0.7, 0.7, 1.0}

\newcommand\nodata{ ~$\cdots$~ }%

\title{The high-speed X-ray camera on AXIS}

\author[a]{Eric D.\ Miller}
\author[a]{Marshall W.\ Bautz}
\author[a]{Catherine E.\ Grant}
\author[a]{Richard F.\ Foster}
\author[a]{Beverly LaMarr}
\author[a]{Andrew Malonis}
\author[a]{Gregory Prigozhin}
\author[a]{Benjamin Schneider}
\author[b]{Christopher Leitz}
\author[c]{Sven Herrmann}
\author[c,d,e]{Steven W.\ Allen}
\author[c]{Tanmoy Chattopadhyay}
\author[c]{Peter Orel}
\author[c,d]{R.\ Glenn Morris}
\author[c,e]{Haley Stueber}
\author[f]{Abraham D.\ Falcone}
\author[g]{Andrew Ptak}
\author[h]{Christopher Reynolds}
\author[h]{the AXIS Team}

\affil[a]{Kavli Institute for Astrophysics and Space Research, Massachusetts Institute of Technology, Cambridge, MA, USA}
\affil[b]{Lincoln Laboratory, Massachusetts Institute of Technology, Lexington, MA, USA}
\affil[c]{Kavli Institute for Particle Astrophysics and Cosmology, Stanford University, Stanford, CA, USA}
\affil[d]{SLAC National Accelerator Laboratory, 2575 Sand Hill Road, Menlo Park, CA, USA}
\affil[e]{Department of Physics, Stanford University, 382 Via Pueblo Mall, Stanford, CA, USA}
\affil[f]{Department of Astronomy and Astrophysics, Pennsylvania State University, University Park, PA, USA}
\affil[g]{NASA Goddard Space Flight Center, Greenbelt, MD, USA}
\affil[h]{Department of Astronomy, University of Maryland, College Park, MD, USA}

\authorinfo{Further author information:\\E.~D.~Miller: E-mail: milleric@mit.edu}

\begin{document} 
\maketitle

\begin{abstract}
AXIS is a Probe-class mission concept that will provide high-throughput, high-spatial-resolution X-ray spectral imaging, enabling transformative studies of high-energy astrophysical phenomena. To take advantage of the advanced optics and avoid photon pile-up, the AXIS focal plane requires detectors with readout rates at least 20 times faster than previous soft X-ray imaging spectrometers flying aboard missions such as Chandra and Suzaku, while retaining the low noise, excellent spectral performance, and low power requirements of those instruments. We present the design of the AXIS high-speed X-ray camera, which baselines large-format MIT Lincoln Laboratory CCDs employing low-noise pJFET output amplifiers and a single-layer polysilicon gate structure that allows fast, low-power clocking. These detectors are combined with an integrated high-speed, low-noise ASIC readout chip from Stanford University that provides better performance than conventional discrete solutions at a fraction of their power consumption and footprint. Our complementary front-end electronics concept employs state of the art digital video waveform capture and advanced signal processing to deliver low noise at high speed. We review the current performance of this technology, highlighting recent improvements on prototype devices that achieve excellent noise characteristics at the required readout rate. We present measurements of the CCD spectral response across the AXIS energy band, augmenting lab measurements with detector simulations that help us understand sources of charge loss and evaluate the quality of the CCD backside passivation technique. We show that our technology is on a path that will meet our requirements and enable AXIS to achieve world-class science.
\end{abstract}


\section{Introduction}
\label{sect:intro}

The Advanced X-ray Imaging Satellite (AXIS)\cite{AXIS,Reynoldsetal2023} is a mission concept in response to the Astrophysics Probe Explorer (APEX) call, providing high-throughput, high-spatial-resolution X-ray imaging spectroscopy. The unique capabilities of AXIS will allow ground-breaking studies addressing key science priority areas identified by the National Academies’ 2020 Decadal Survey on Astronomy and Astrophysics\cite{Astro2020}. In particular, AXIS will explore ``Cosmic Ecosystems'' by studying the birth and evolution of super-massive black holes and the mechanisms of galactic feedback; it will help place ``Worlds and Suns in Context'' by studying the affects of stellar activity on the planets they harbor; and it will be a key player in ``New Messengers and New Physics'' thanks to a rapid on-board transient alert system and rapid response time to external transient triggers. AXIS will provide the crucial X-ray counterpart to the panchromatic suite of large observatories of the 2030s, including JWST, Rubin, Roman, LIGO/Virgo/Kagra, LISA, SKA, and Euclid.

The high throughput and spatial resolution of AXIS create technical challenges for the detector system. While similar instruments have flown with great success on missions such as Chandra and Suzaku, to avoid photon pile-up, AXIS requires a camera operating at least 20 times faster than on those heritage missions. At the same time, to allow ground-breaking astrophysics studies, the AXIS detectors must retain or exceed the excellent spectral imaging performance of Chandra ACIS and Suzaku XIS. As we have shown in previous work, in some cases these requirements are at odds with each other technically. For example, AXIS requires excellent quantum efficiency and spectral response across a wide 0.2--10 keV energy band. To accomplish this at the hard end requires a detector at least 100 $\mu$m thick, yet such a thick silicon detector with pixels small enough to sample the PSF can struggle to meet the soft X-ray performance requirement due to diffusion of charge from photons interacting far from the collection gates\cite{Milleretal2018,Milleretal2022c}. The AXIS camera design follows careful consideration of the baseline mission performance, shown in Table \ref{tab:axis_reqs} along with the derived camera requirements.

\begin{table}[t]
\caption{AXIS baseline parameters relevant to the camera. \label{tab:axis_reqs}}
\footnotesize
\begin{center}       
\begin{tabular}{|l|l|} 
\hline\hline
\multicolumn{2}{|l|}{\textbf{AXIS mission parameters}} \\ \hline
Spatial resolution at 1 keV (HPD) & 1.25\arcsec\ (on-axis) \\
                              & 1.50\arcsec\ (FoV-average) \\\hline 
Effective area at 1 keV       & 4200 cm$^2$ (on-axis) \\
                              & 3600 cm$^2$ (FoV-average) \\\hline
Effective area at 6 keV       & 830 cm$^2$ (on-axis) \\
                              & 570 cm$^2$ (FoV-average) \\\hline
Field of view                 & 24\arcmin\ diameter \\\hline
Energy band                   & 0.2--10 keV \\\hline
Energy resolution (FWHM)      & $\leq$70 eV (at 1 keV) \\
                              & $\leq$150 eV (at 6 keV)\\\hline
Orbit                         & circular low-Earth orbit \\
                              & $i<8$\arcdeg, 610--680 km \\ \hline
Prime mission lifetime              & 5 years\\ \hline
\multicolumn{2}{|l|}{\textbf{Focal Plane Assembly characteristics}} \\ \hline
Frame rate                  & $\geq$ 5 fps (goal 20 fps)\\ \hline
Serial readout rate         & $\geq$ 2 MHz \\ \hline
Pixel size                  & 24 $\mu$m (0.55\arcsec) \\ \hline
Readout noise               & $\leq$ 3 e- RMS \\ \hline
Focal plane temperature     & $-90 \pm 0.1$\arcdeg C \\ \hline
\hline
\end{tabular}
\end{center}
\end{table} 

In this contribution, we describe the design for the AXIS camera, called the Focal Plane Assembly (FPA). We also review recent developments of our focal plane performance, which continue a multi-year effort to develop fast, low-noise detectors for future strategic X-ray missions.\cite{Bautzetal2019,Prigozhinetal2020,Lamarretal2020, Bautzetal2020, LaMarretal2022, Prigozhinetal2022, Milleretal2022c, LaMarretal2022b, Bautzetal2022,Herrmannetal2020,Chattopadhyayetal2020,Oreletal2022,Herrmannetal2022,Chattopadhyay2022_ccd,Chattopadhyayetal2022,Chattopadhyay2023_sisero}. We finally demonstrate that the advanced technology is on track to reach the required technical readiness level for the AXIS mission.

\section{AXIS Camera Design}
\label{sect:design}

The high-speed AXIS X-ray camera incorporates a focal plane array of fast-readout charge-coupled devices (CCDs) designed and fabricated by MIT Lincoln Laboratory (MIT/LL) and building on a long line of successful space instruments spanning the last three decades. Each CCD is coupled with an application-specific integrated circuit (ASIC) specifically designed by Stanford University to provide low-noise and low-power amplification of the CCD analog signal. The front-end electronics incorporate modern digital processing to further reduce noise. The back-end electronics under development at Penn State implement a state-of-the-art FPGA-based Event Recognition Processor (ERP) to greatly reduce the telemetry stream, and they also include a Transient Alert Module (TAM) to detect changes in flux among sources in the field of view and rapidly disseminate transient alerts to the community. The detector system is housed in a reliable, high-heritage camera structure that builds on lessons learned from previous missions.

\subsection{Detector system}

At the heart of the camera are four MIT/LL CCD detectors arranged in a 2$\times$2 array to cover the 24\arcmin\ AXIS field of view. Each frame-store CCD is back-illuminated for enhanced soft X-ray sensitivity, 100 $\mu$m thick for hard X-ray sensitivity, and has 24-$\mu$m (0.55\arcsec) pixels to sample the sharp AXIS PSF. The AXIS CCID-100 detector builds on heritage devices such as the CCID-41 back-illuminated device that flew as XIS1 on Suzaku, sharing several design features including pixel size and charge injection implementation. Two key technical advancements allow faster operation without a loss of performance or increase in power consumption. First, the triple layer of polysilicon used for the clocking gate structures in the CCID-41 and similar devices has been replaced with a single polysilicon layer\cite{}. This allows the gates to be located very near each other, in turn requiring much lower voltage swings during charge transfer and thus much lower power consumption per transfer. Second, the single-stage on-chip MOSFET output amplifier has been updated to a two-stage pJFET amplifier, producing similar noise levels at $\sim$ 10 times faster readout rate\cite{}. Each CCD has eight outputs to increase the data rate from the 1440$\times$1440 pixel imaging area. The CCD backside is passivated with a molecular beam epitaxy (MBE) process that deposits a thin 5--10 nm layer of heavily doped silicon. Performance results obtained by our group at the MIT Kavli Institute (MKI) from prototype CCDs with these design features have been reported over the past several years\cite{Bautzetal2019,Prigozhinetal2020,Lamarretal2020, Bautzetal2020, LaMarretal2022, Prigozhinetal2022, Milleretal2022c, LaMarretal2022b, Bautzetal2022}, and updates are presented in Section \ref{sect:ccdperf}.

Traditional off-chip amplification done with discrete components is not suitable for our purposes due to the amount of power required to reduce the parasitic capacitance and the real-estate these components would occupy. Our group at Stanford University has developed an ASIC called the Multi-Channel Readout Chip (MCRC) specifically for use with these MIT/LL CCDs. The MCRC is designed in a 350-nm technology node featuring 8 channels. Each channel has two selectable gain settings, an input referred noise of 1.63 e$^{-}$ RMS, an input dynamic range of $\pm$320 mV, channel-to-channel crosstalk less than $-75$ dBc, a power consumption of roughly 31 mW/channel, and a bandwidth of approximately 50 MHz, translating to an effective rise time of around 5 ns. Such a response can comfortably support readout speeds for large CCD pixel matrices in excess of 5 Mpixel/s/channel. The ASIC also features 8 integrated current sources to bias the CCD outputs. The heart of the MCRC is a fully differential amplifier with a switched capacitive feedback to minimize added noise. Along with being optimized for low noise it also provides two gain settings of 6 and 12 V/V. The amplifier is configured such that it translates the input single-ended CCD signal to a fully differential output. The signal is then buffered to the outside of the ASIC by a fully differential unity-gain amplifier designed to drive a transmission line to support digital waveform sampling with the commercial ADCs that we deploy in the Front End Electronics.  The MCRC is currently deployed and functioning in test systems at MIT and Stanford, as described in Section \ref{sect:asicperf}.

\begin{figure}[t]
\begin{center}
\includegraphics[width=.67\linewidth]{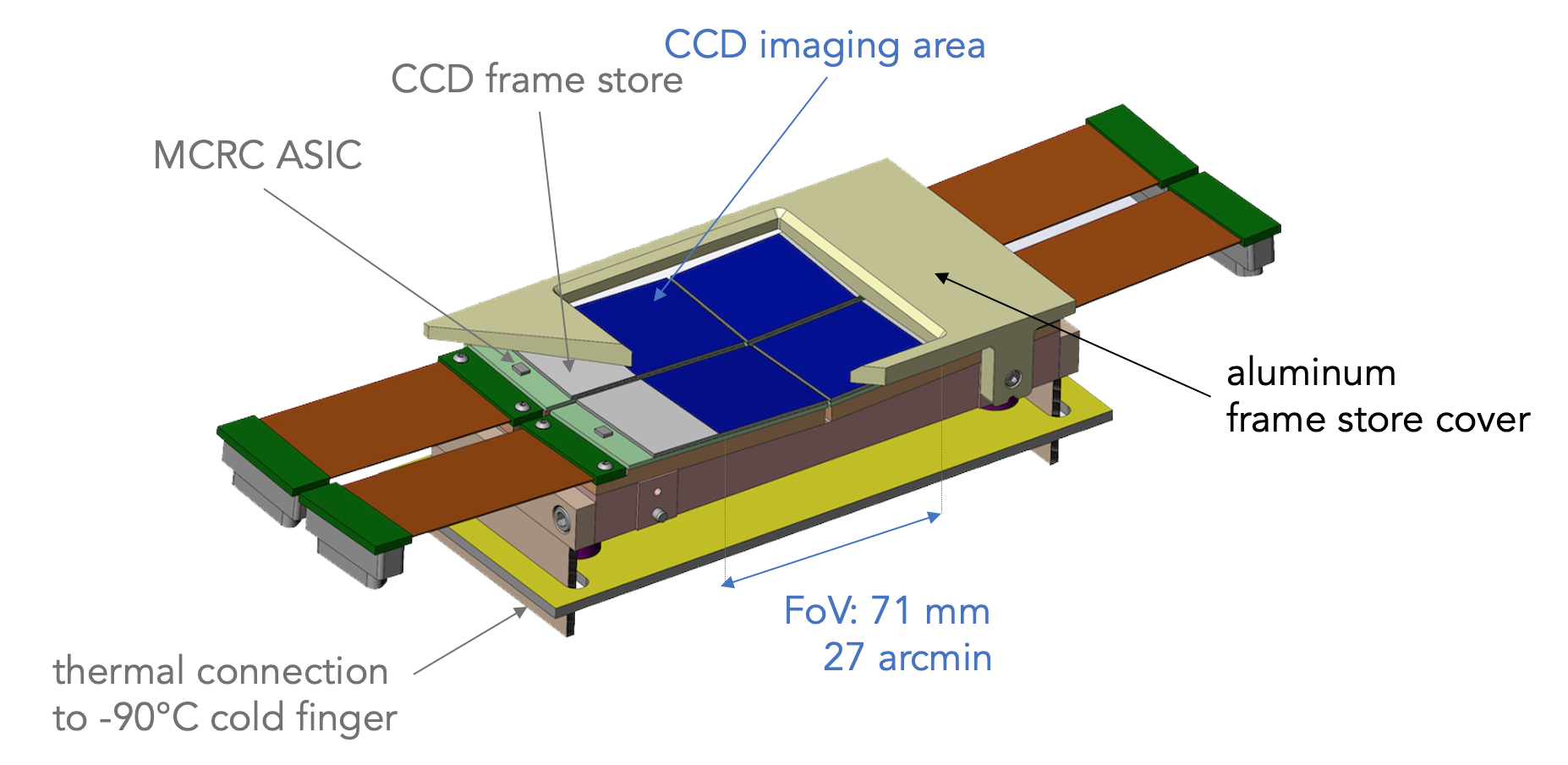}
\end{center}
\caption{AXIS detector array with components labeled. The array consists of four fast-readout frame-store CCDs arranged in a 2$\times$2 pattern, each with a dedicated ASIC readout chip co-mounted on the interposer. An aluminum cover shields the frame store regions from focused celestial X-rays during readout, and it provides the ASICs with additional particle damage protection. The entire array is maintained at -90\arcdeg C using a cold finger connection and active trim heaters.}
\label{fig:detarray}
\end{figure}

Each CCD and ASIC pair is mounted on the same detector package, shown in Figure \ref{fig:detarray}, and operates as a unit. For eight CCD outputs and ASIC channels per detector running at 2 MHz, and a parallel transfer speed of 1 MHz, the current best estimate for CCID-100 frame rate is 7 frames per second (fps), meeting the AXIS requirement of 5 fps. This configuration results in 0.5\% out-of-time fraction, and so the frame rate could be increased without requiring faster parallel transfer. Increasing the frame rate is possible through design changes such as increasing the number of outputs; or through operational changes, such as increasing the output rate to 5 MHz. This latter change may come at the cost of increased noise and reduced soft X-ray response, and so could be implemented as a choice based on the science goal. Similarly, time resolution for point-like sources could be improved to the sub-ms regime by reading out only a small sub-array of the aimpoint detector or even employing a continuous-clocking mode that eliminates all spatial information along CCD columns. We are confident that we can meet the AXIS goal of 20 fps with some combination of design and operational considerations, while retaining good spectral performance to allow ground-breaking science. Because of this, all electronics downstream of the detector system have been designed to accommodate this faster pixel rate.

\subsection{Mechanical design}

\begin{figure}[p]
\begin{center}
\includegraphics[width=.4\linewidth]{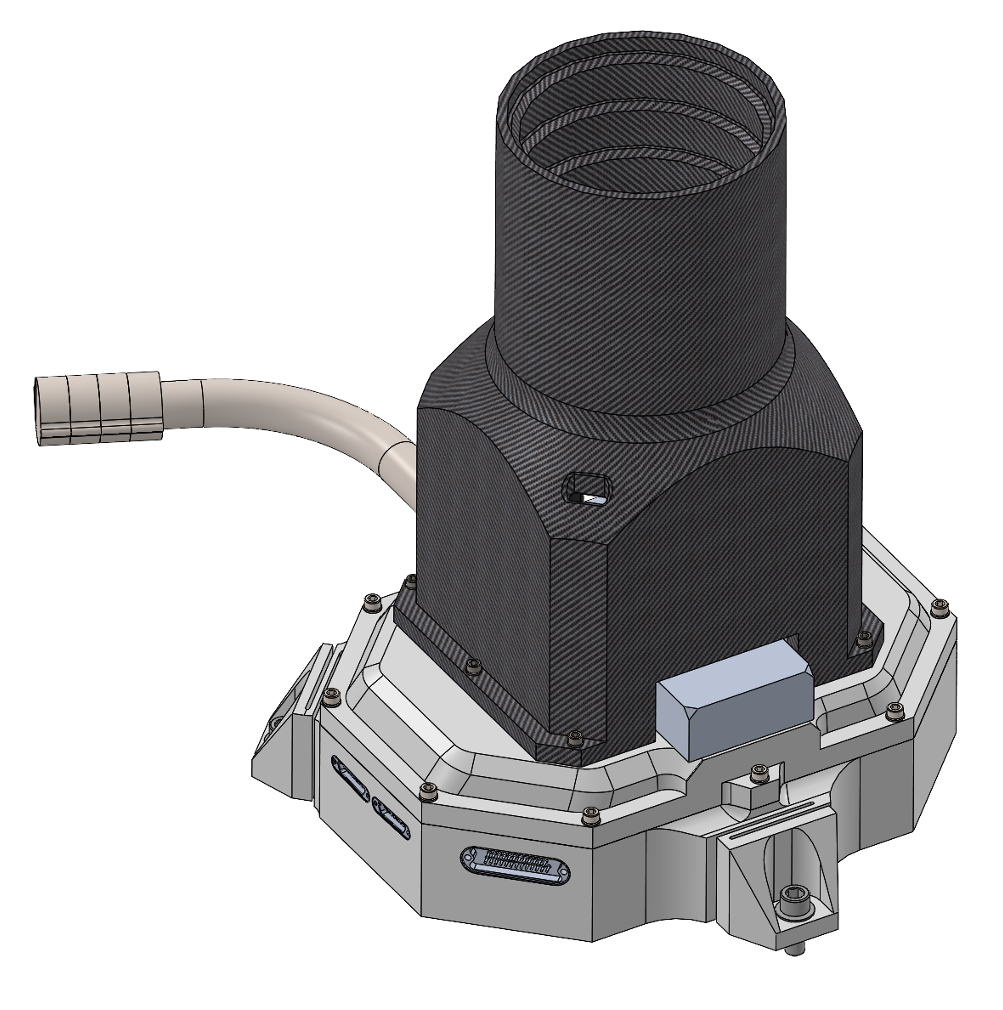}
\hspace*{.5in}
\includegraphics[width=.35\linewidth]{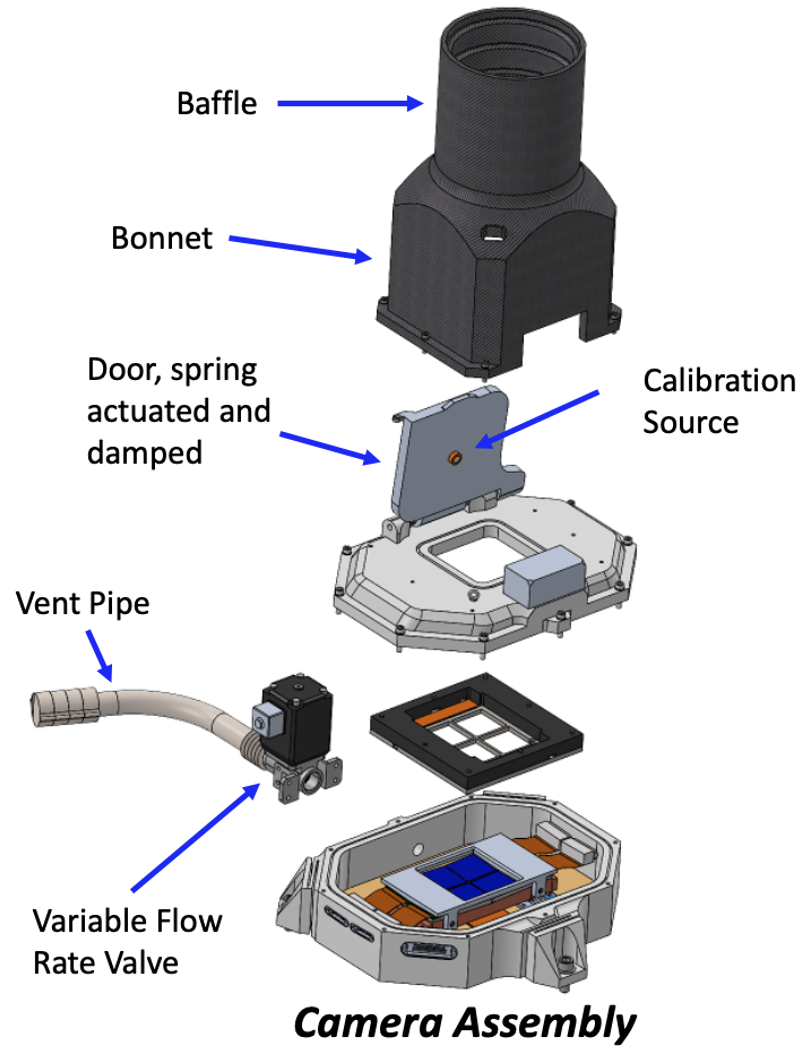}
\end{center}
\caption{The AXIS focal plane camera assembly, rendered as integrated (left) and in an exploded view (right). Various features are labeled and described further in the text.}
\label{fig:housing}
\end{figure}

The focal plane is housed in a vacuum enclosure that incorporates high-heritage components and lessons learned from previous missions. This housing, shown in Figure \ref{fig:housing} with its various components, is kept under vacuum on the ground to reduce the risk of molecular contamination build-up. The detector array is thermally isolated from the housing with standoffs, and connected to a cold finger that cools the detectors and ASICs to $-90$\arcdeg C, eliminating dark current and mitigating the effects of particle-induced radiation damage. This temperature is maintained to $\pm$0.1\arcdeg C using trim heaters. The vacuum is maintained with a commandable vent valve and a one-time-open door; in orbit the vent valve is permanently opened to the vacuum of space. A warm contamination blocking filter serves the dual purpose of blocking optical and UV light and, along with proper contamination control at all stages of assembly and integration, preventing the build-up of molecular contamination along the light path. Similar contamination compromised the soft X-ray performance of instruments aboard Chandra and Suzaku\cite{}. A bonnet and baffle provide structural support and stray-light blocking, and radioactive $^{55}$Fe sources mounted on the closed door and illuminating unused corners of the CCDs provide continuous, well-characterized reference photons of $\sim$ 6 keV for detector monitoring and calibration.

\begin{figure}[p]
\begin{center}
\includegraphics[width=\linewidth]{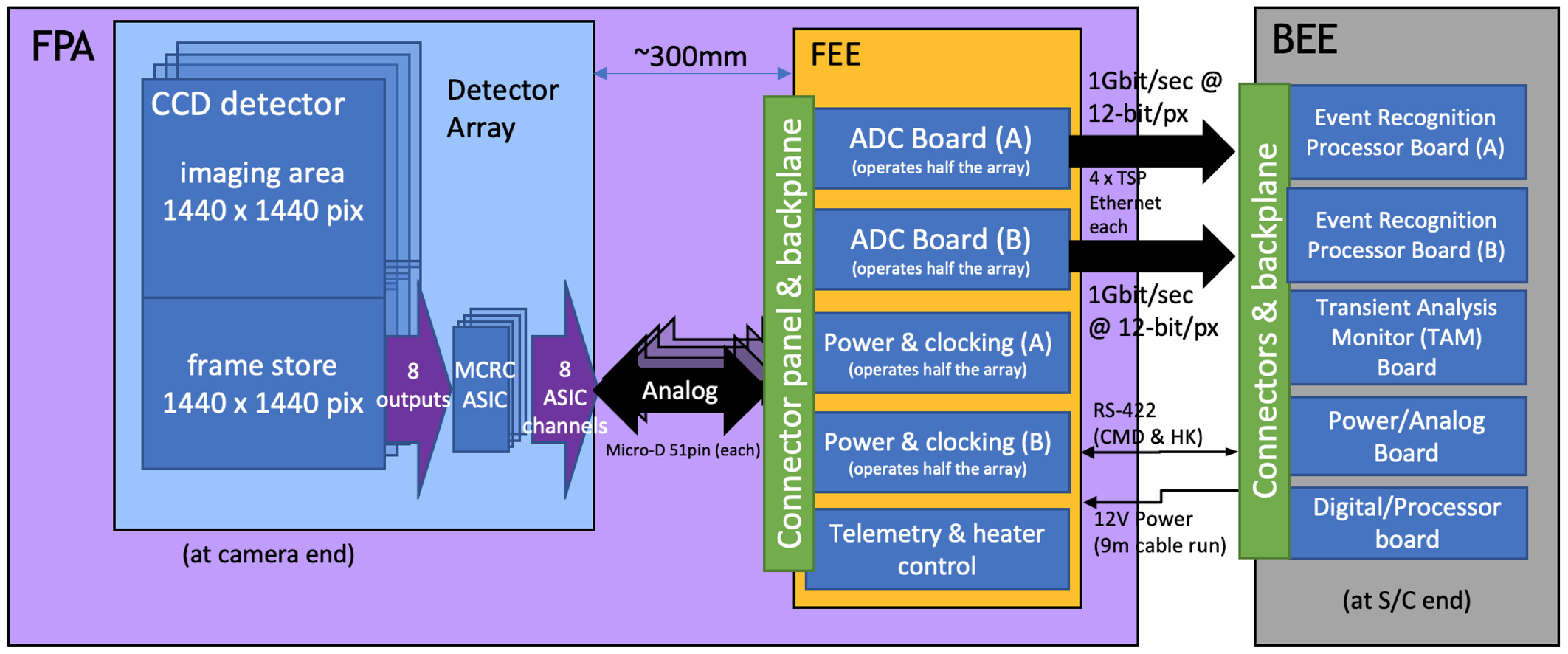}
\end{center}
\caption{Block diagram of the FPA and BEE showing components and interfaces between the sub-systems. Within the FEE, the (A) and (B) indicate boards dedicated to a pair of detectors. Within the BEE, the (A) and (B) Event Recognition Processor (ERP) boards are fully redundant thanks to cross-strapping; each ERP board will nominally read pixel data from a single FEE ADC board (a pair of detectors), but a single ERP board can receive and process data from the entire focal plane, even at the goal frame rate of 20 fps.}
\label{fig:blockdiagram}
\end{figure} 

\subsection{Front End Electronics}
\label{sect:fee}

The CCD+ASIC detectors are controlled and sampled by the Front End Electronics (FEE), which share heritage with several highly successful past missions, including ASCA, Chandra, Suzaku, TESS, and eROSITA. The interfaces between the detectors and the FEE, and further downstream between the FEE and Back End Electronics (BEE), are shown in Figure \ref{fig:blockdiagram}. The FEE contains five 6U printed circuit boards connected to a backplane within an electronics box enclosure that provides the connector panel interface. Two of these boards, the Power and Clocking Boards, provide fully differential bias and sequenced clock voltages to the CCDs as well as bias voltages to the readout ASICs. One board operates a pair of detectors independently from the other board. The ADC Boards receive the detector video signal from the ASIC outputs and amplify and digitize the pixel voltages, again with one board responsible for a pair of detectors. These FEE boards are highly complementary to the CCD+ASIC detector system, employing Microchip PolarFire FPGAs and ST Microelectronics ADCs to perform state-of-the-art digital video waveform capture at 40 Msamples/s, delivering low noise at high speed. The fifth Telemetry and Heater Control board controls all thermal and mechanical components in the camera assembly, including the trim heaters that maintain the focal plane temperature, the heaters keeping the contamination blocking filter warm, and single actuation of the camera vent valve and door. This board also receives commands and relays HK data along a dedicated LVDS line to the BEE. Digitized pixel data is transferred from the ADC Boards to the BEE along four dedicated Ethernet lines that utilize a Microchip gigabit Ethernet PHY layer. To minimize noise and impedance loss along the analog signal line, the FEE is located within 1 m of the camera.

\subsection{Back End Electronics}
\label{sect:bee}

The BEE receives an image of digitized pixel values for each CCD frame from the FEE and processes this into a list of candidate X-ray events. This processing is performed by a pair of FPGA-based Event Recognition Processor (ERP)\cite{Burrowsetal2016} boards, which first correct the images for pixel-to-pixel and time-dependent bias offsets and mask bad pixels using maps held in memory. Local maxima are identified as candidate events along with the neighboring 3$\times$3 pixels that exceed tunable noise thresholds. Information about the location, summed pulse height, and pixel pattern or ``grade'' are extracted for each candidate event, which are then filtered based on a configurable set of rules and packaged for telemetry. Each ERP board carries half the load simultaneously; however, all CCD lines are strapped to both boards, thus enabling intrinsic redundancy with a single ERP board capable of processing data from all detectors, even at the goal frame rate of 20 fps.

A copy of the event stream is sent to the Transient Analysis Module (TAM) within the BEE to search for transient sources and enable alerts to be issued in real time. The TAM software is based on that currently flying on the Swift observatory and also on algorithms initially developed for the Athena Science Products Module\cite{Pradhanetal2020}. It compares changes in detected sources with an on-board source catalog and with its own recent history, accounting for pointing uncertainties and effects such as dithering. Transient alerts meeting a configurable set of criteria are transferred to the ground and disseminated to the community within 10 minutes\cite{Reynoldsetal2023}.

\section{Current performance of the detector system}

\subsection{CCD performance}
\label{sect:ccdperf}

\subsubsection{Summary of CCD types and MKI test facilities}
\label{sect:ccdsetup}

\begin{figure}[p]
\begin{center}
\includegraphics[height=2in]{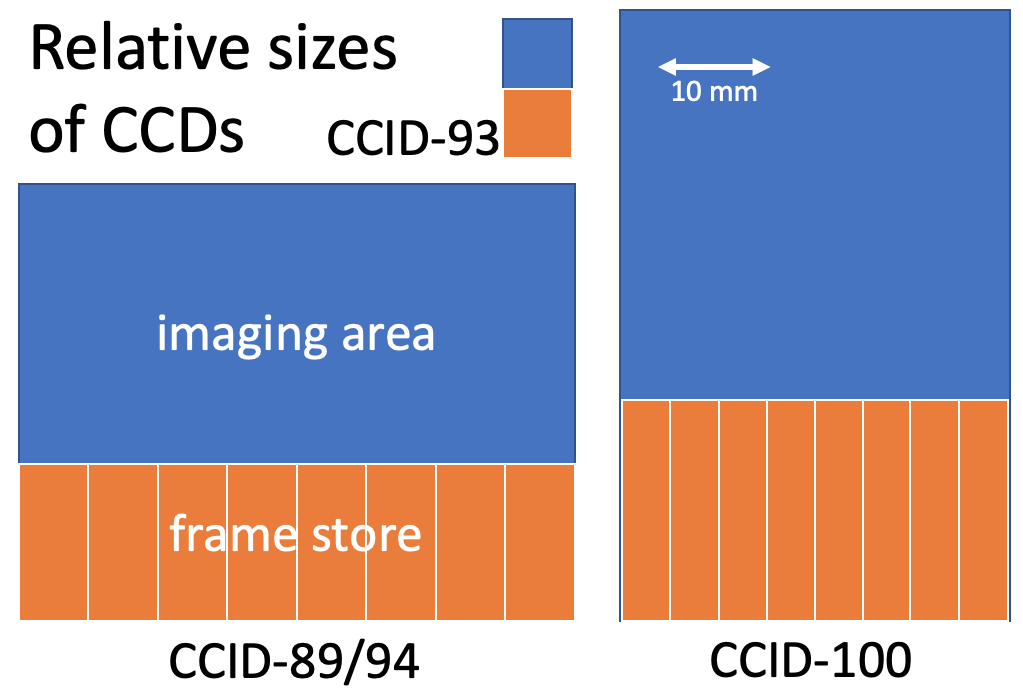}
\hspace*{.2in}
\includegraphics[height=2.5in]{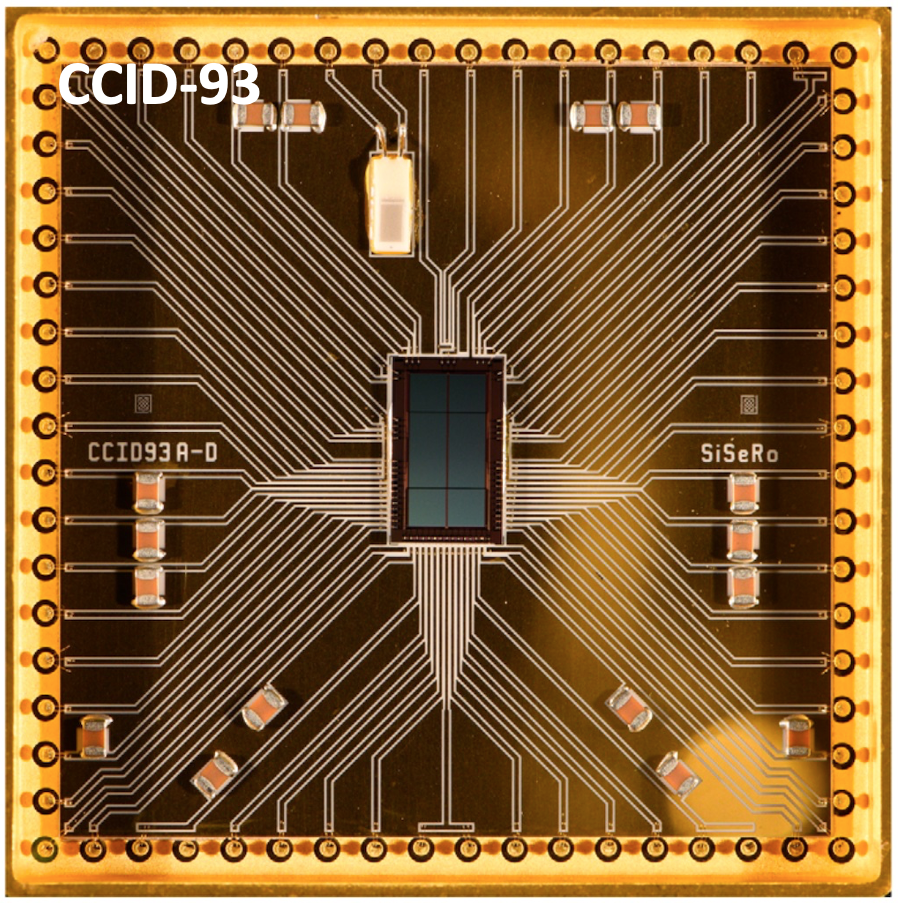}\\
\vspace*{.2in}
\includegraphics[height=3in]{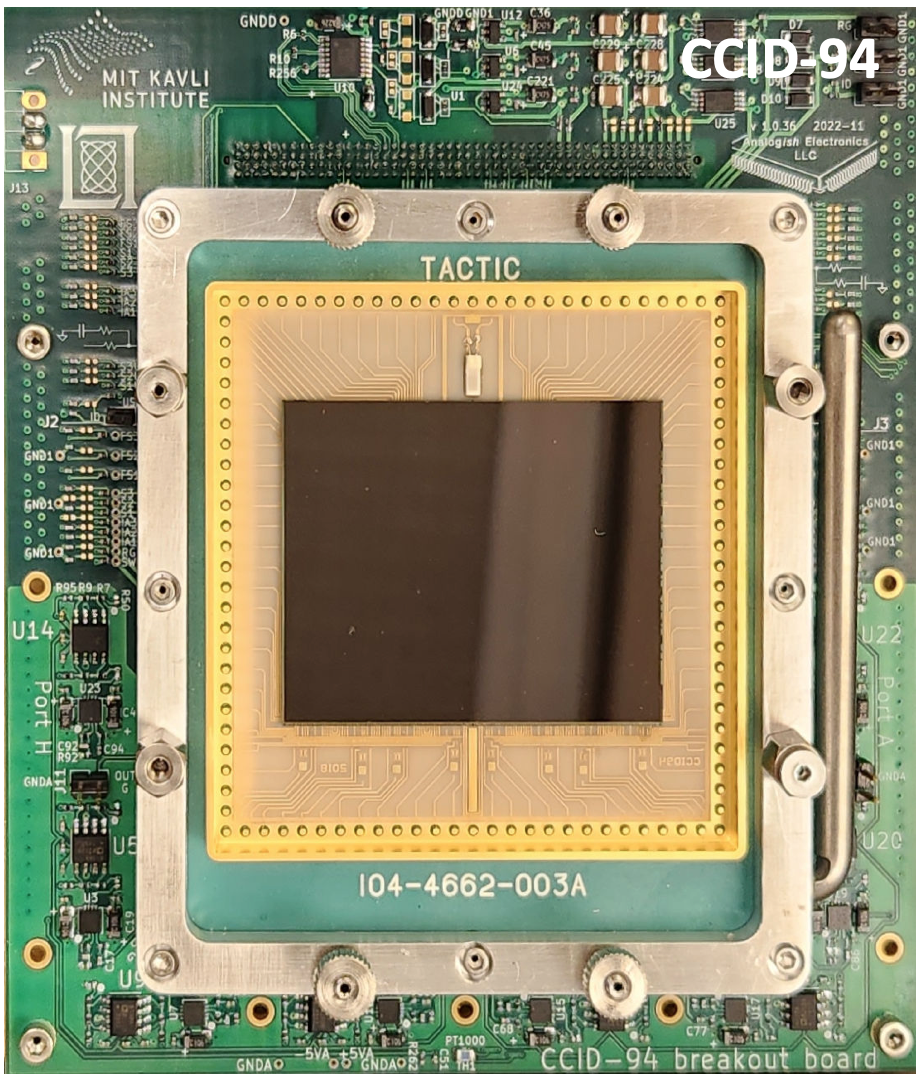}
\hspace*{.2in}
\includegraphics[height=3in]{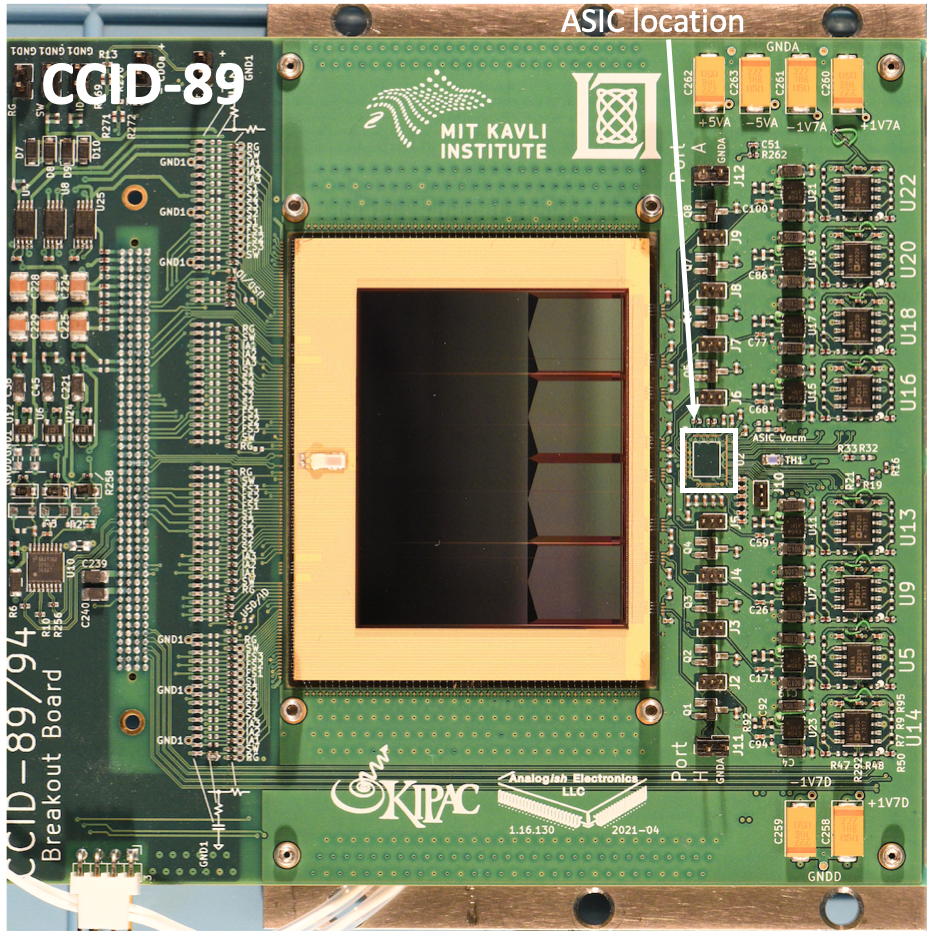}
\end{center}
\caption{Photographs of packaged MIT/LL CCDs under testing for AXIS development and performance demonstration. Photos are not at the same scale; the schematic in the upper left panel shows the actual relative sizes and the number of output nodes as orange segments in the frame store region. While the CCID-94 and -89 are the same size, they are shown in different orientations. The CCID-89 shown is a front-illuminated device, so the metal layer is visibly delineating the frame store regions, which cannot be seen in the back-illuminated CCID-94 shown here. The CCID-100 is the AXIS design CCD and has not been fabricated, so there is no photo.}
\label{fig:protoccds}
\end{figure}

Our groups at MIT/LL and MKI have continued development of advanced, fast-readout CCD detectors for a strategic mission such as AXIS for the past several years. Here we provide an update on the performance of various CCD types, building on results presented in previous work.\cite{Bautzetal2019,Prigozhinetal2020,Lamarretal2020, Bautzetal2020, LaMarretal2022, Prigozhinetal2022, Milleretal2022c, LaMarretal2022b, Bautzetal2022} These CCDs are shown in their test packages in Figure \ref{fig:protoccds}, and their features are summarized in Table \ref{tab:protoccds} along with the current AXIS CCD design. While the CCID-93 and CCID-94 are useful devices for characterizing certain features of the AXIS CCID-100, the CCID-89 is identical in most ways to that expected future device, and can be considered the ``prototype'' AXIS detector. 

The test facilities at MKI include dedicated vacuum chambers for each of the CCD types, each equipped with a liquid-nitrogen cryostat for thermal control to temperatures below $-100$\arcdeg C. Each setup includes an Archon\footnote{\url{http://www.sta-inc.net/archon}} controller that provides CCD bias and clock voltages and performs digital sampling on the CCD analog video waveform, incorporating an interface board that connects to a custom made vacuum feed-through board and detector board for each CCD. The lab detector package and boards for each CCD were designed by MIT/LL. Each chamber allows easy insertion of a radioactive $^{55}$Fe source for reliable full-frame illumination with Mn K$\alpha$ (5.9 keV) and K$\beta$ (6.4 keV) X-rays. The CCDID-89 and -94 setups can incorporate a $^{210}$Po source with Teflon target that produces fluorescence lines of C K (0.53 keV) and F K (0.68 keV) across the full 50$\times$25 mm imaging area. For the testing presented here, the CCID-93 chamber is mounted on an In-Focus Monochromator (IFM)\cite{Hettrick1990_ifm} that uses grazing incidence reflection gratings to produce clean monochromatic lines at energies below 2 keV, with typical spectral resolving power $\lambda/\Delta\lambda = E/\Delta E \sim $ 60--80, far higher than that of the CCD itself. We are currently adapting the IFM setup to allow testing with the CCID-89 and -94 chambers.

All test data is acquired as a set of full pixel frames that are processed in a similar way to flight data on Chandra and to the expected BEE algorithm (see Section \ref{sect:bee}). Briefly, each image is corrected for pixel-to-pixel and time-dependent bias levels. Persistent (``hot'') pixels are masked, and local maxima above a defined event threshold (5--8 times the RMS noise) are identified. Pixel islands around these maxima are searched for pixels above a second ``split'' threshold (3--4 times the RMS noise), and all such pixels are summed to produce the event pulse height, a measurement of the incident photon energy. This summed pulse height is denoted ``allaboveph'' in several figures in this work. Pixel islands are typically 7$\times$7 for the CCID-93 with 8-$\mu$m pixels, and 3$\times$3 for the others devices with 24-$\mu$m pixels. Each event is assigned a ``multiplicity'', akin to the ``grade'' on ASCA, Chandra, and Suzaku, but here simply encoding the number of pixels ``n\#'' that are above the split threshold.

Where appropriate, we include detector simulations in an effort to understand the detector physics and its effect on performance. These simulations use \texttt{Poisson CCD}\cite{Lageetal2021}, a software package that models the 3-D electric field in a specified silicon detector and performs a Monte Carlo simulation of the drift and diffusion of photoelectrons introduced into the detector volume. The simulation setup is similar to what we have presented in past work\cite{Milleretal2018,LaMarretal2022,Milleretal2022c}.

\begin{table}[t]
\caption{Features of MIT/LL CCDs under testing for AXIS at MKI and Stanford, compared to the AXIS CCD design.}\label{tab:protoccds}
\begin{center}       
\scriptsize
\begin{tabular}{|p{1in}|>{\centering}p{1.2in}|>{\centering}p{1.12in}|>{\centering}p{1.12in}|>{\centering\arraybackslash}p{1.35in}|}
\hline\hline
\textbf{Feature}                   & \textbf{CCID-93} & \textbf{CCID-94} & \textbf{CCID-89} & \textbf{CCID-100 (AXIS)} \\ \hline
Format                    & Frame-transfer, 512$\times$512 pixel imaging array                       & \multicolumn{2}{|c|}{Frame-transfer, 2048$\times$1024 pixel imaging array} & Frame-transfer, 1440$\times$1440 pixel imaging array \\ \hline
Image area pixel size  & 8$\times$8 $\mu$m                                                            & 24$\times$24 $\mu$m                & 24$\times$24 $\mu$m             & 24$\times$24 \\ \hline
Output ports              & 1 pJFET, 1 SiSeRO (independent)                                     & 8 pJFET                   & 8 pJFET                & 8--16 pJFET \\ \hline
Transfer gate design      & Single layer polysilicon                                            & Triple layer polysilicon  & Single layer polysilicon & Single layer polysilicon \\ \hline
Additional features       & Regions with 0.5, 1, 2 $\mu$m and no trough; charge injection & Trough, charge injection          & Trough, charge injection       & Trough, charge injection\\ \hline
BI detector thickness     & 50 $\mu$m                                                               & 50 $\mu$m                     & 50 $\mu$m                  & 100 $\mu$m \\ \hline
Back surface              & MIT/LL MBE 5--10 nm; JPL delta doping                               & MIT/LL MBE 10 nm          & MIT/LL MBE 5--10 nm    & MIT/LL MBE 5--10 \\ \hline
Typical serial rate       & 2--5 MHz                                                           & 0.5 MHz                   & 2--5 MHz              & $\geq$2 MHz \\ \hline
Typical parallel rate     & 0.1 MHz                                                             & 0.1 MHz                   & 0.2 MHz                & $\geq$0.5 MHz\\ \hline
Full frame read time   & 0.15--0.56 s                                                       & 1 s                       & 0.15--0.56 s          & $\leq$0.2 s \\
\hline\hline
\end{tabular}
\end{center}
\end{table} 

\subsubsection{CCD X-ray performance}

The ability of a detector to accurately report the energy of an incoming photon is crucial to X-ray imaging spectroscopy. This is one of the key challenges for AXIS due to its sensitivity and broad energy band, as we have explained in Section \ref{sect:intro} and elsewhere\cite{LaMarretal2022,Milleretal2022c}. In addition to Fano noise, a variety of effects can degrade the detector response to photons of different energies, including signal lost below the noise threshold; signal lost to charge transfer inefficiency (CTI) or other charge traps\cite{LaMarretal2022b}; signal lost to defects at the backside entrance window; and signal redistributed to other energies due to fluorescence ionization within the detector. 

\begin{figure}[t]
\begin{center}
\includegraphics[width=.49\linewidth]{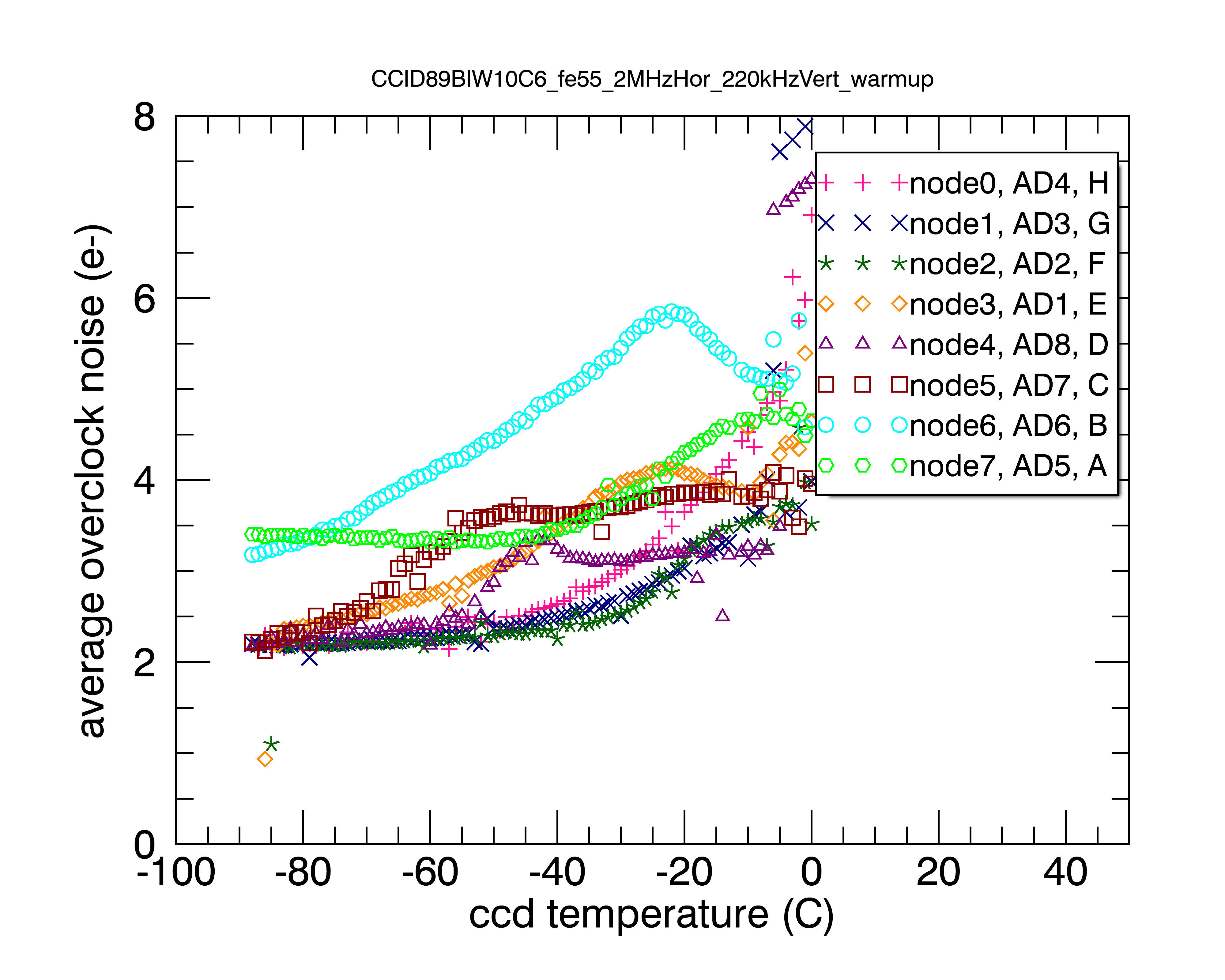}
\includegraphics[width=.47\linewidth]{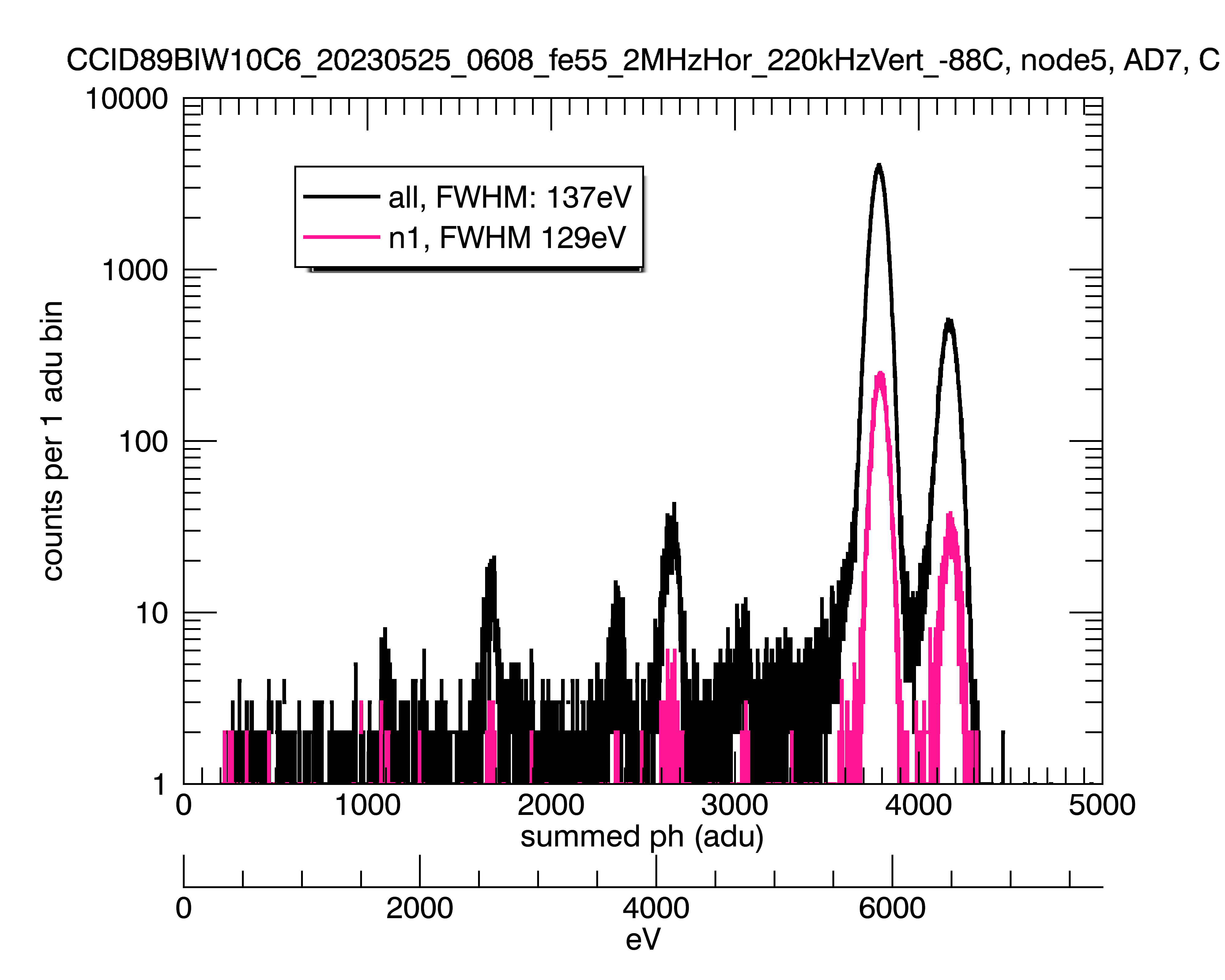}
\end{center}
\caption{(left) Readout noise as a function of temperature for all pJEFT nodes on the back-illuminated CCID-89 (W10C6). At this readout rate of 2 MHz, the AXIS baseline CCID-100 will reach 7 fps frame rate. Six of the eight nodes have lower noise than the required 3 e- RMS at the AXIS operating temperature of $-90$\arcdeg C. (right) Spectrum from a representative node of the same CCID-89 produced under illumination by $^{55}$Fe. The bright lines of Mn K$\alpha$ (5.9 keV) and K$\beta$ (6.4 keV) are easily resolved and show that we meet the AXIS requirements for spectral resolution at 6 keV ($\leq$150 eV FWHM), even combining all events of different pixel multiplicities.}
\label{fig:89noisefe55}
\end{figure}

Minimizing readout noise is a key challenge for the AXIS detector in order to achieve excellent spectral performance, since any pixel-to-pixel stochastic noise such as that introduced by the readout chain produces an irreducible broadening of the response profile. Our latest testing of a back-illuminated CCID-89 (identified as W10C6) shows that the pJFET outputs perform remarkably well at 2 MHz at a variety of temperatures (see Figure \ref{fig:89noisefe55}). This readout rate would provide a frame rate of 7 fps for the AXIS CCID-100, meeting our requirement. At the AXIS operating temperature of $-90$\arcdeg C, the noise is less than 2.5 e- RMS for six of the eight output, meeting our requirement. These measurements were taken with modest tuning of clock settings to achieve the best noise for a single output. With additional tuning, we anticipate meeting the noise requirement for all nodes. Further noise measurements from our test CCDs are presented in Section \ref{sect:su93} and elsewhere in these proceedings.\cite{Prigozhinetal2023}.

The measured spectral response bears out the excellent noise results. In Figure \ref{fig:89noisefe55}, we also show a spectrum of $^{55}$Fe from a single representative segment of the same CCID-89. Mn K$\alpha$ and K$\beta$ are seen near 6 keV, along with escape and fluorescence lines at lower energies. The Mn K$\alpha$ line profile is Gaussian over more than two orders of magnitude, with a width of 137 eV FWHM including events of all pixel multiplicities. This performance exceeds the AXIS requirement of 150 eV FWHM. Similar results are seen for other segments on this device, and for other well-performing segments of our other test detectors.

The response below 1 keV is key for the study of high-redshift objects and low temperature X-ray-emitting plasma. In a back-illuminated CCD, soft response depends on the backside passivation quality, since soft X-rays interact close to the entrance window. In addition, since the interaction site is furthest away from the collection gates, charge diffusion is maximized, and signal (which is smaller to begin with) is more likely to spread to multiple pixels and be lost below the noise threshold. Understanding the importance of each effect is valuable.

\begin{figure}[p]
\begin{center}
\includegraphics[width=.49\linewidth]{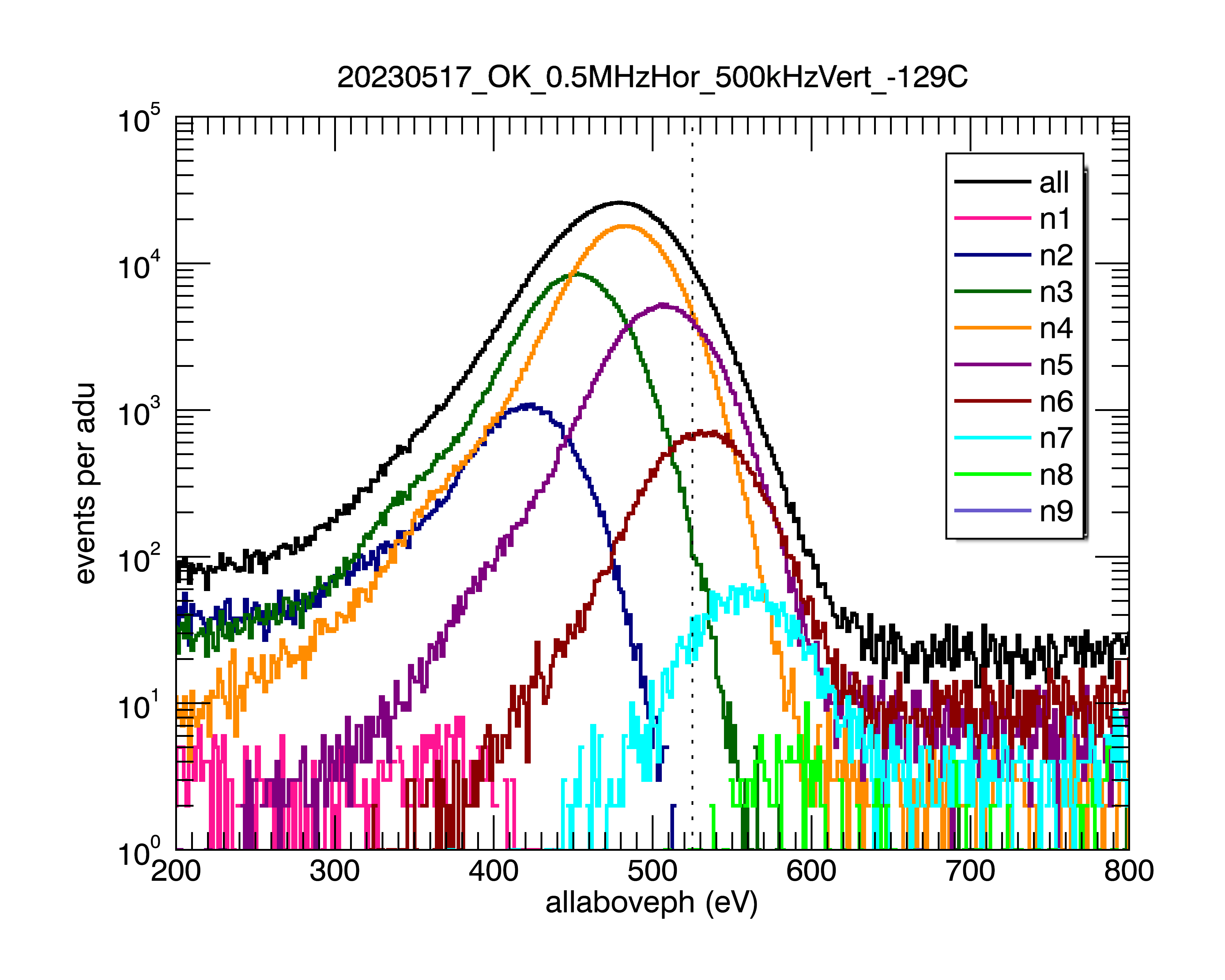}
\includegraphics[width=.47\linewidth]{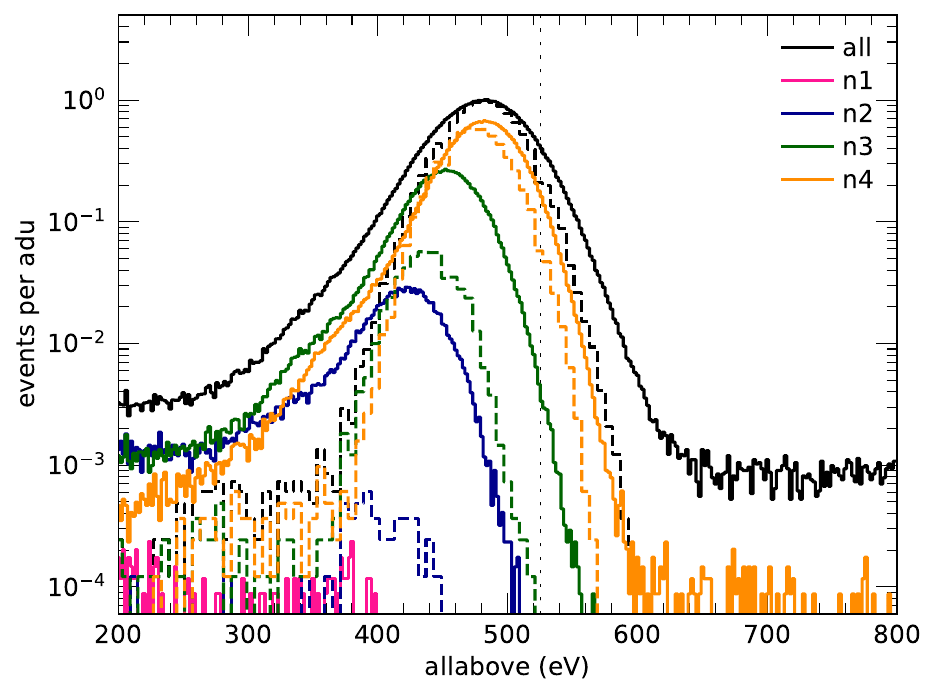}
\end{center}
\caption{(left) Spectrum of the O K fluorescence line from the CCID-93 illuminated by the IFM. Spectra from events with different pixel multiplicities are plotted separately. ``allaboveph'' in the X axis indicates the event energies were estimated by summing the pulse heights of all pixels above the noise threshold. Details of the processing and further interpretation can be found in the text. (right) Simulated spectra for the same detector and lab setup (dashed lines), plotted with a selection of spectra from the left panel (solid lines). The simulations include charge diffusion and the effects of enforcing a noise threshold, but do not reproduce the low-energy tail, indicating a different source of charge loss. The multiplicity distribution is also different.}
\label{fig:93ifm}
\end{figure}

We have characterized the soft X-ray response in back-illuminated versions of all three CCD models. Due to its smaller 8-$\mu$m pixels, the CCID-93 is invaluable to study the effects of charge diffusion to separate it from other sources of signal loss, such as entrance window passivation quality. We illuminated this detector with O K 0.53 keV photons using the IFM, and the resulting spectra are shown in Figure \ref{fig:93ifm} (left panel). The readout speed of 0.5 MHz was selected to achieve the lowest noise (1.7 e- RMS) in order to identify non-noise-related charge loss effects. Spectra from events with different pixel multiplicities are plotted separately; there is a clear energy shift for different multiplicities, although the widths of the dominant multiplicities are all less than 70 eV FWHM (see Table \ref{tab:MKIresults}). In principle, a multiplicity-dependent gain correction\cite{Schneider2023_svom_mxt} can significantly improve the overall FWHM. Non-Gaussian low-energy tails indicate some source of charge loss. Spectra from simulations of same detector and lab setup are shown in the right panel of Figure \ref{fig:93ifm} along with a selection of spectra from the left panel. The simulations include charge diffusion and the effects of enforcing a noise threshold, but do not reproduce the low-energy tail, indicating a different source of charge loss. The multiplicity distribution is also different for the lab data and simulations. The source of the soft tail remains under investigation, however it is clear that even with small pixels, the MBE backside processing allows us to achieve the soft spectral response required for AXIS.

\begin{table}[p]
\caption{Results from CCD testing at MKI.}\label{tab:MKIresults}
\begin{center}       
\scriptsize
\begin{tabular}{|ll|c|c|c|}
\hline\hline
\multicolumn{2}{|l|}{\textbf{Typical result}} & \textbf{CCID-93} & \textbf{CCID-94} & \textbf{CCID-89} \\ \hline
\multicolumn{2}{|l|}{Detector temperature}    & $-128$\arcdeg       &  $-89$\arcdeg      & $-87$\arcdeg C  \\ \hline
\multicolumn{2}{|l|}{Serial readout rate}     &  0.5 MHz     &  0.5 MHz     & 2 MHz        \\ \hline
\multicolumn{2}{|l|}{Readout noise (RMS)}     &  1.7 e-    &  2.1--3.1 e- &  2.2--3.4 e- \\ \hline
\multicolumn{5}{|l|}{Spectral resolution (Gaussian FWHM)$^a$} \\ \hline
C K & all     & \nodata          &  \nodata       &  74 eV (100\%)\\ 
0.27 keV      & n1          & \nodata          &  \nodata       &  71 eV (48\%) \\ 
              & n2          & \nodata          &  \nodata       &  75 eV (45\%) \\ 
              & n3          & \nodata          &  \nodata       &  90 eV (6\%) \\
\hline
O K & all     & 83 (100\%)       &  \nodata       &  \nodata\\ 
0.53 keV      & n2          & 66 (4\%)         &  \nodata       &  \nodata\\ 
              & n3          & 63 (25\%)        &  \nodata       &  \nodata\\ 
              & n4          & 63 (53\%)        &  \nodata       &  \nodata\\ 
              & n5          & 65 (16\%)        &  \nodata       &  \nodata\\ 
              & n6          & 67 (2\%)         &  \nodata       &  \nodata\\
\hline
F K           & all         & \nodata          &  66 eV (100\%) &  76 eV (100\%) \\ 
0.68 keV      & n1          & \nodata          &  63 eV (36\%)  &  70 eV (21\%)  \\
              & n2          & \nodata          &  66 eV (49\%)  &  75 eV (52\%) \\
              & n3          & \nodata          &  80 eV (9\%)   &  80 eV (19\%) \\
              & n4          & \nodata          &  74 eV (6\%)   &  81 eV (7\%) \\
\hline
Mn K          & all          & 142 eV (100\%)   & 139 eV (100\%) & 137 eV (100\%) \\
5.9 keV       & n1           & 123 eV (5\%)     & 129 eV (18\%)  & 129 eV (6\%) \\
              & n2           & \nodata          & 136 eV (44\%)  &  \nodata \\
              & n3           & \nodata          & 140 eV (18\%)  &  \nodata \\
              & n4           & \nodata          & 143 eV (20\%)  &  \nodata \\
\hline
\hline
\multicolumn{5}{l}{$^a$Where available, spectral resolution is provided for each pixel multiplicity} \\
\multicolumn{5}{l}{~`n\#', with the fraction of events with that multiplicity given in parentheses.} \\
\multicolumn{5}{l}{~Missing data indicates either lack of a measurement at that energy or lack of} \\
\multicolumn{5}{l}{~multiplicity information in the analysis. Multiplicities representing fewer} \\
\multicolumn{5}{l}{~than 1\% of events are also not shown.} \\
\end{tabular}
\end{center}
\end{table} 

\afterpage{\clearpage}

\begin{figure}[t]
\begin{center}
\includegraphics[width=.90\linewidth]{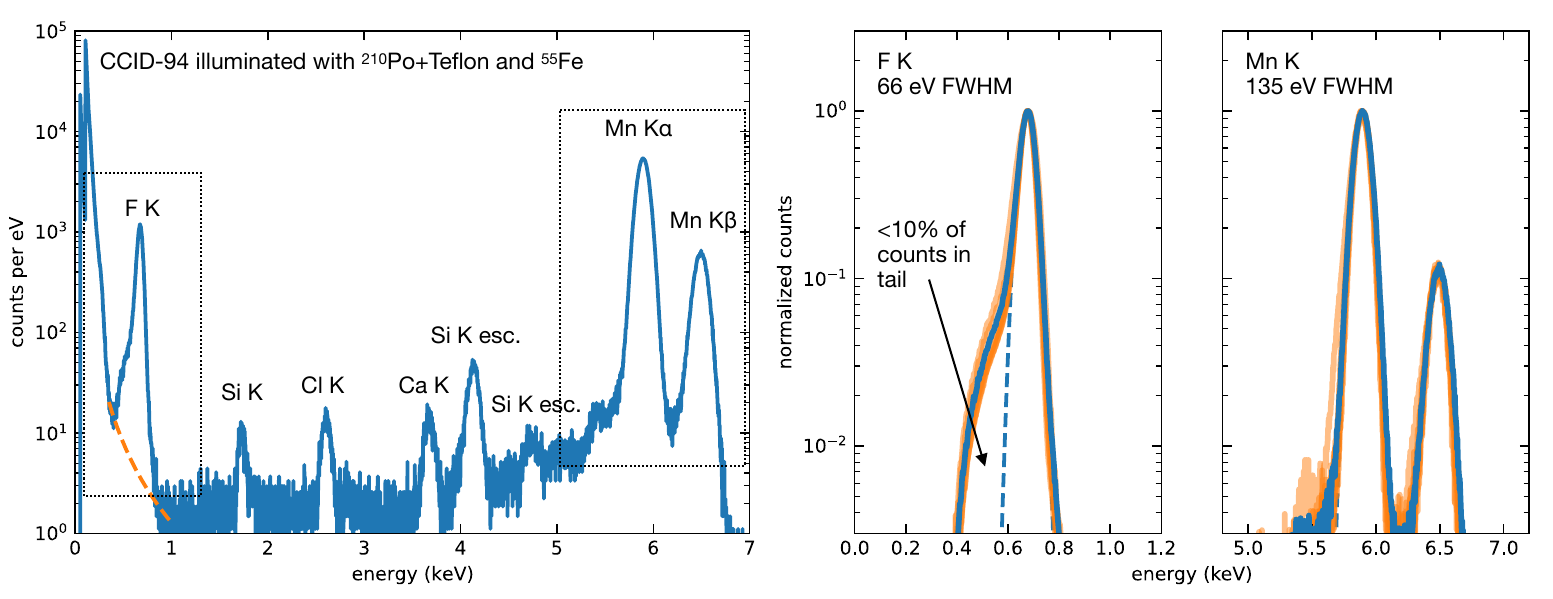}
\end{center}
\caption{(left panel) Spectrum of a single CCID-94 segment (node `C') simultaneously illuminated with $^{210}$Po with a Teflon target and $^{55}$Fe. The primary fluorescence lines of F K, Mn K$\alpha$, and Mn K$\beta$ can be seen along with several other fluorescence and escape features. (right two panels) Zoom-in of the F K and Mn K peaks from the left panel, now also showing spectra from the other seven nodes of this chip as orange curves. The F K line has been corrected for the noise continuum by subtracting a best-fit power law, shown in dashed orange in the left panel. Gaussian fits (dashed blue lines) indicate that spectral FWHM meets the AXIS requirement at both energies. While the F K peak shows a non-Gaussian tail, it contains fewer than 10\% of the line counts.}
\label{fig:94pofe55}
\end{figure}

The CCID-94 is also useful to understand the soft response. We illuminated our test device with $^{210}$Po with Teflon target and $^{55}$Fe simultaneously to produce a broad-band spectrum from a single representative segment (see Figure \ref{fig:94pofe55}).  The primary fluorescence lines of F K, Mn K$\alpha$, and Mn K$\beta$ can be seen along with several other fluorescence and escape features. C K is not visible due to a low-energy noise peak that is unrelated to the detector or source. After fitting and subtracting a power-law model to this noise continuum, we measure the spectral resolution at F K (0.68 keV) to be 66 eV FWHM for all event multiplicities for this segment, with little variation from segment to segment. There is a non-Gaussian tail, which remains under investigation but could result from backside surface losses. This tail contains less than 10\% of the counts for all segments, and its segment-to-segment uniformity indicates it can be accounted for using a standard redistribution matrix. The response at $\sim$ 6 keV is excellent for all nodes, averaging 135 eV FWHM for all event multiplicities. Results are summarized in Table \ref{tab:MKIresults}.

\begin{figure}[t]
\begin{center}
\includegraphics[width=.46\linewidth]{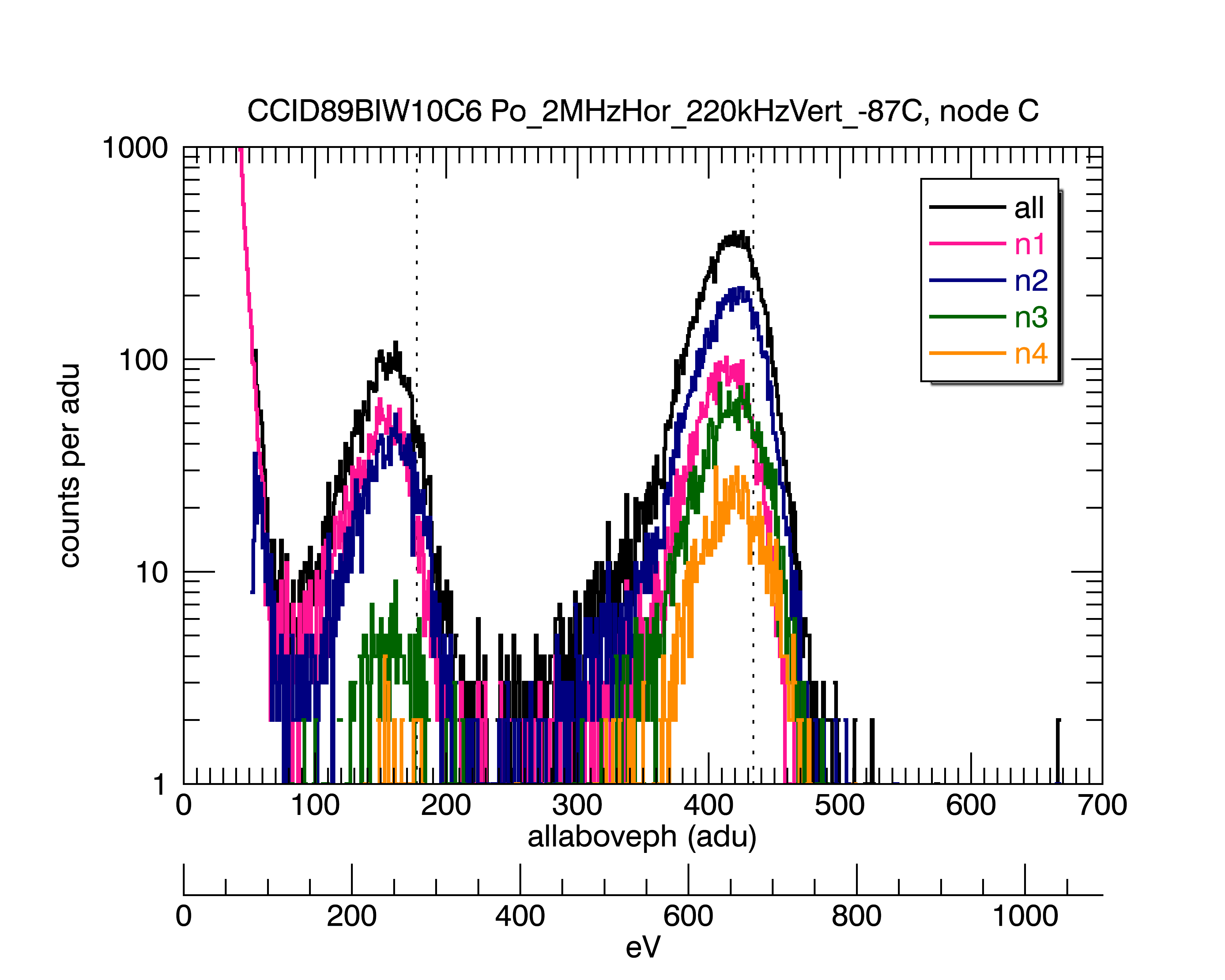}
\includegraphics[width=.47\linewidth]{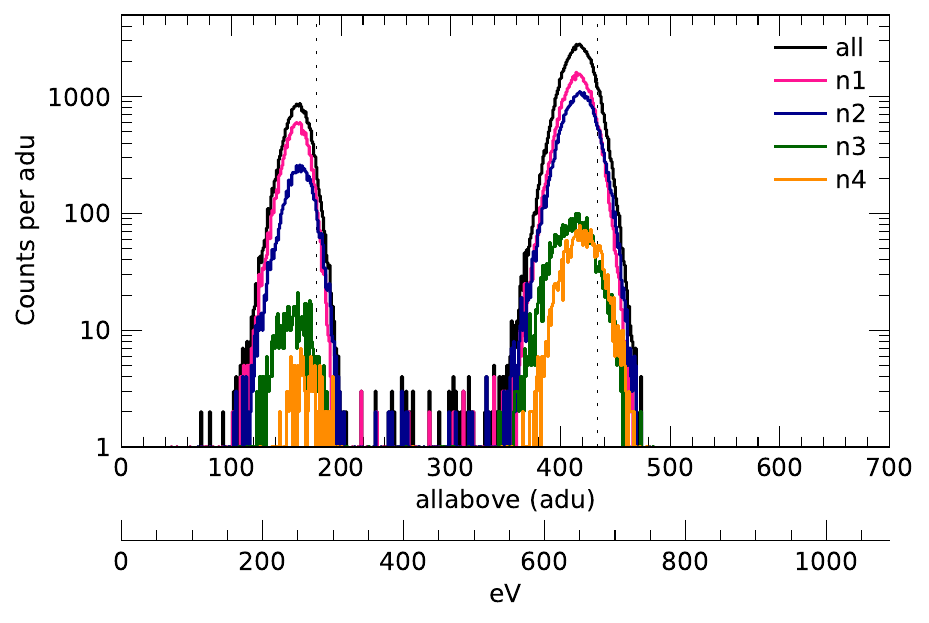}
\end{center}
\caption{(left) Spectrum of the C K and F K fluorescence lines from the CCID-89 illuminated by a $^{210}$Po source with a Teflon target. This spectrum is from a representative CCD segment, and spectra from events with different pixel multiplicities are plotted separately. (right) Spectral lines produced in a simulation of the same detector and lab setup show considerably narrower FWHM and somewhat different multiplicity distributions.}
\label{fig:89po}
\end{figure} 

The CCID-89 operating at 2 MHz is the closest approximation of the future AXIS CCD, featuring a similar structure and including all effects of charge collection and fast charge transfer that can alter the soft X-ray response. A spectrum from $^{210}$Po with Teflon target illuminating a representative segment of this chip is shown in Figure \ref{fig:89po}. We clearly detect both the C K (0.28 keV) and F K (0.68 keV) from the Teflon and easily separate them from the noise peak. The spectral FWHM, shown in Table \ref{tab:MKIresults}, approaches the AXIS requirement of 70 eV FWHM below 1 keV, although it is wider than simulations of charge diffusion and noise threshold effects would predict (Figure \ref{fig:89po}, right panel).

All three CCD flavors exhibit low noise and excellent X-ray response across the AXIS band and, for the multi-output models, across multiple segments. While the soft response formally meets AXIS requirements, each detector shows some disagreement with simulations, generally producing broader FWHM with a tail toward low energies. This is indicative of a charge loss mechanism not included in simulations, such as CTI or (more likely) charge losses at the backside entrance window. We are planning to test the CCID-89 and CCID-94 with the IFM soon, and this will allow illumination with cleaner monochromactic lines at more soft energies, such as C K (0.27 keV), O K (0.53 keV), and Mg K (1.25 keV). It should also eliminate the noise source from the current CCID-93 setup and allow us to probe to lower energies. These results and updated simulations including backside passivation effects will be presented in future work.

\subsubsection{CCID-93 sub-channel implant variant performance}
\label{sect:su93}

\begin{figure}[t]
\centering   
\includegraphics[width=0.55\linewidth]{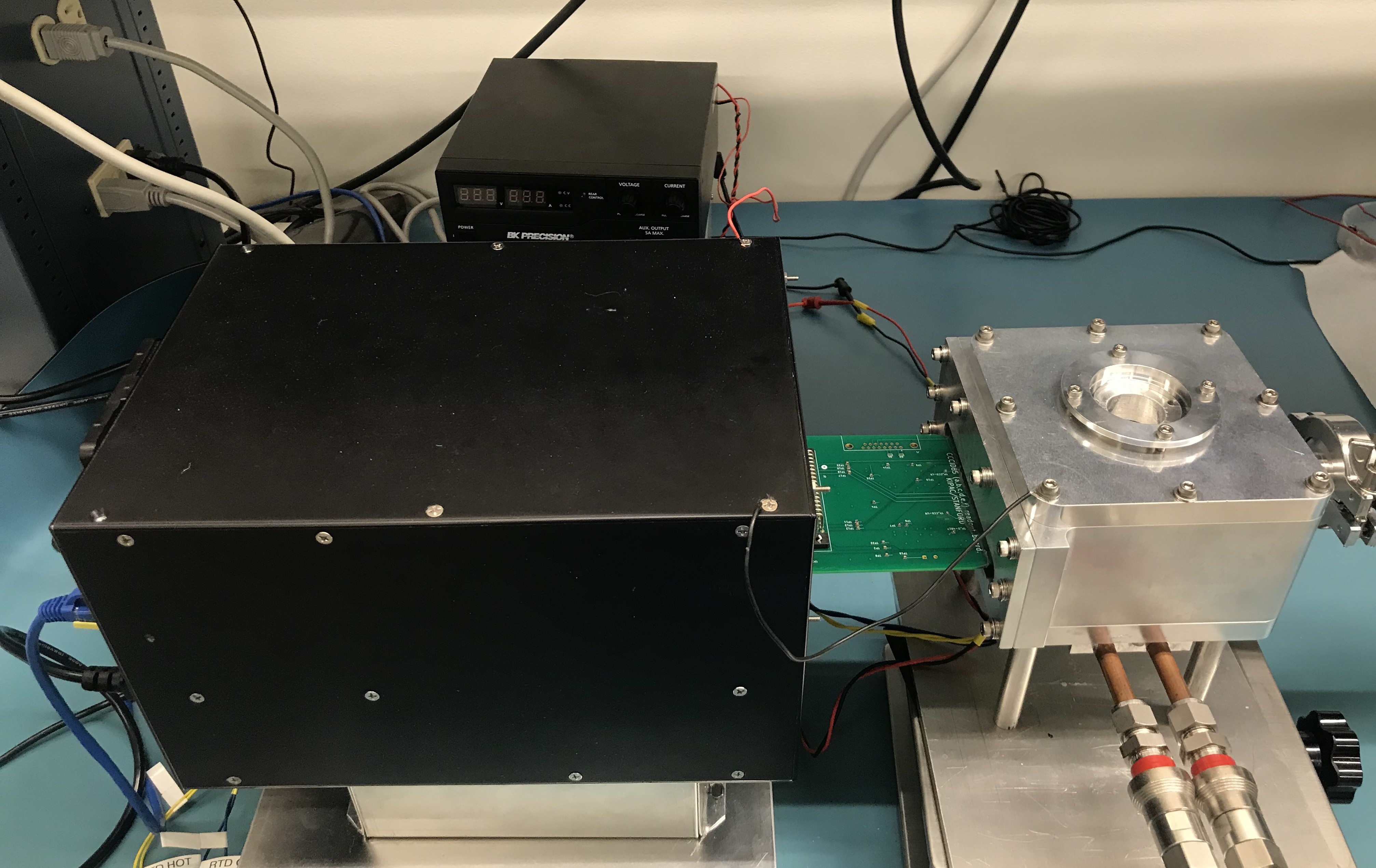}
\hspace*{.1in}
 \includegraphics[width=0.4\linewidth]{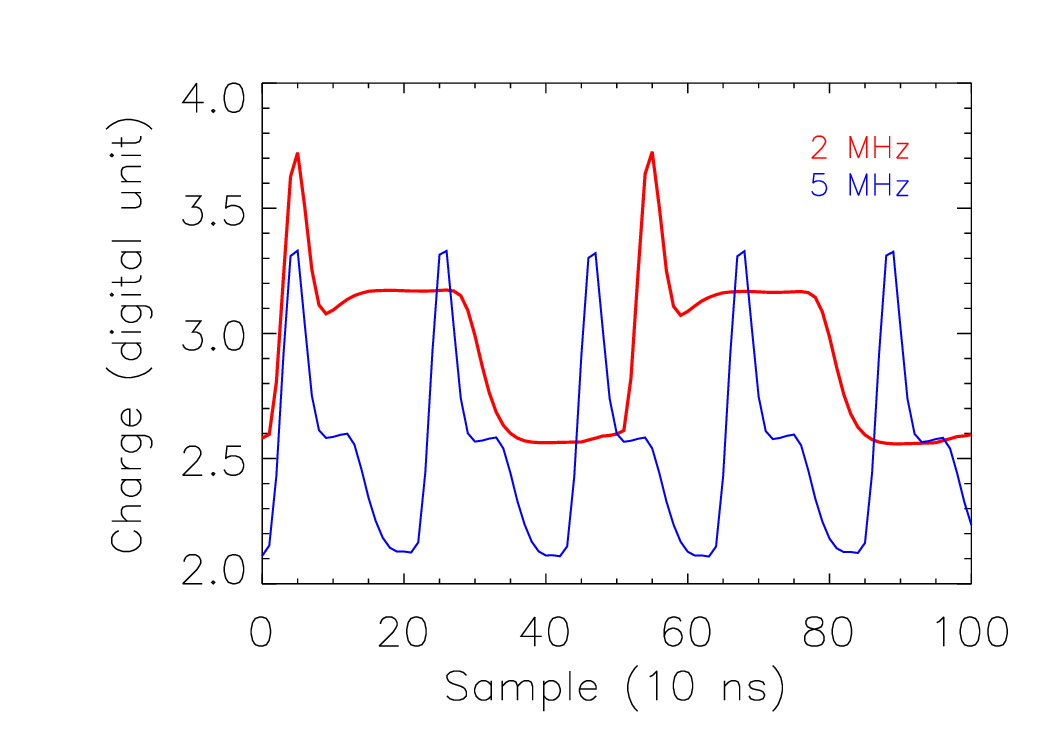} 
\caption{(left) The Stanford University experimental set up with a CCID-93 device mounted inside the ``Tiny Box'' vacuum chamber. A beryllium window mounted on the top flange serves as the X-ray entrance window. (right) CCID-93 video waveform obtained from the Archon controller. The red and blue lines are obtained for 2 and 5 Mpix/s readout speeds, respectively. The baseline and signal samples enable correlated double sampling (CDS) for estimation of charge.}
\label{fig:asic93test}
\end{figure}

Our group at Stanford University has tested one of the MIT/LL CCID-93 detector variants where the output stage follows the same two amplifier architectures, but with an additional sub-channel implant beneath the pJFET amplifier gate.  In this setup, the detector was readout by a readout module consisting of 2-stage amplification stages --- a low noise, low input capacitance, large bandwidth amplifier (ADA4817) followed by a differential ADC driver (AD8138). The bandwidth of the readout is around 50 MHz with noise around 10$-$15 nV/$\sqrt{\mathrm{Hz}}$ (the CCD output stage thermal noise is around 20 nV/$\sqrt{\mathrm{Hz}}$). We use an Archon controller to control and run the CCD clocks and bias voltages, and to digitize the CCD analog output. Figure \ref{fig:asic93test} (left) shows the experiment setup (also knows as the ``Tiny Box" chamber) used to characterize the detectors at Stanford University. The CCD temperature was $-25$\arcdeg C for this experiment.

\begin{table}[p]
\caption{Summary of results from the sub-channel implant variant of the CCID-93 detector at $-25$\arcdeg C.}\label{table:results}
\centering
\begin{tabular}{|c|c|c|}
\hline\hline
\textbf{Readout Speed} & \textbf{Read Noise (e- RMS)} & \textbf{FWHM (eV) at 5.9 keV} \\ \hline
 2 MHz &2.3 & 127 eV \\ \hline
3 MHz & 3.0 &128 eV \\ \hline
4 MHz &3.4 &133 eV \\ \hline
5 MHz & 4.1 &142 eV \\ \hline
\hline
\end{tabular}
\end{table}

\begin{figure}[p]
\centering
\includegraphics[width=0.49\linewidth]{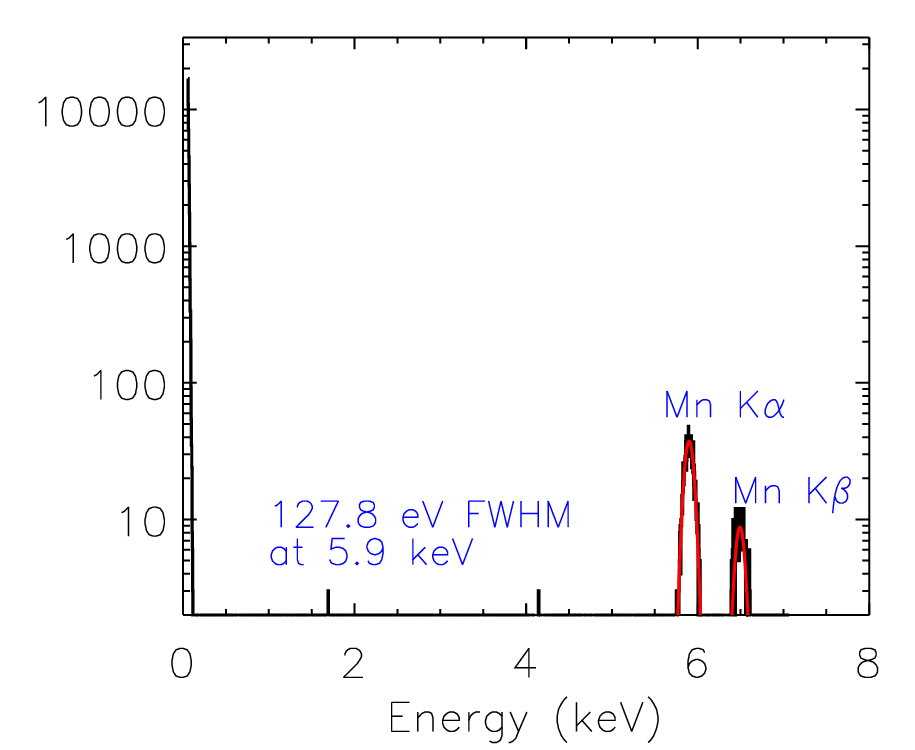} 
\includegraphics[width=0.49\linewidth]{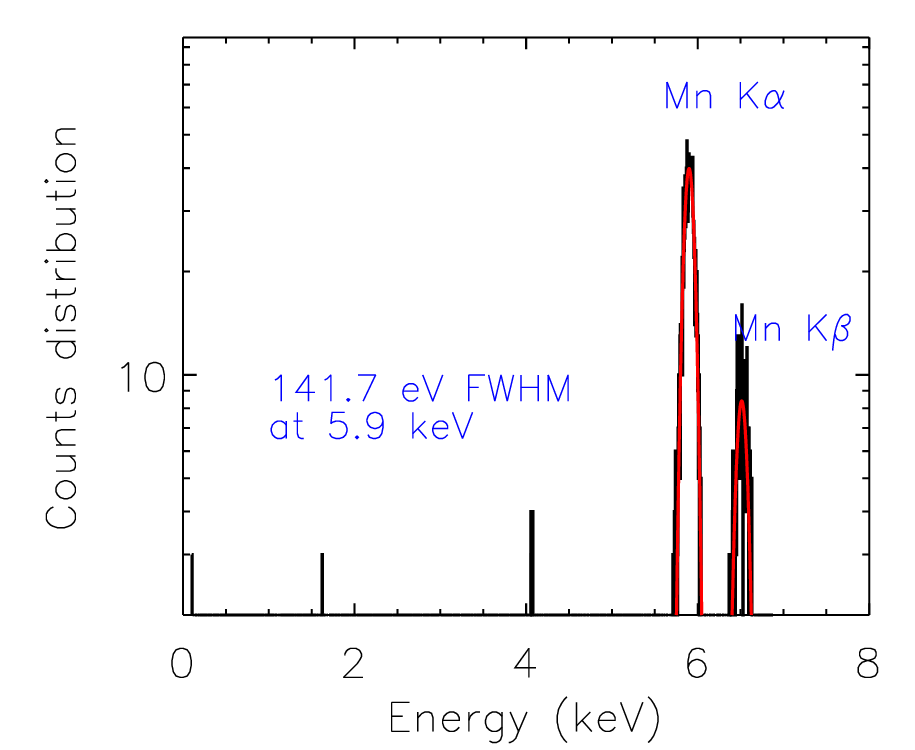}
\caption{Spectrum showing the Mn K$\alpha$ (5.9 keV) and K$\beta$ (6.4 keV) lines from a $^{55}$Fe radioactive source for single-pixel (n0) events for 2 MHz (left) and 5 MHz (right) readout speeds.}
\label{fig:spectra}
\end{figure}

\afterpage{\clearpage}

The detector has been characterized at readout speeds of 2, 3, 4, and 5 Mpix/s. Figure \ref{fig:asic93test} (right panel) shows the CCD digital waveform obtained from the Archon controller for the lowest and highest speeds. As shown in Table \ref{table:results}, the readout noise ranges from 2.3 e- RMS at 2 MHz to 4.1 e- RMS at 5 MHz. For the measured output stage gain of 35 $\mu$V / electron and a transistor transconductance (g$_m$) of 20 $\mu$S, the measured read noise values are very close to the theoretical thermal noise limit of the transistors.  Spectral performance of the device was evaluated using an $^{55}$Fe radioisotope. An X-ray spectrum showing Mn K$\alpha$ (5.9 keV) and Mn K$\beta$ (6.4 keV) lines from the radioactive source is shown in Figure \ref{fig:spectra}. The 5.9 keV line width FWHM is 128 eV for 2 MHz and 142 eV for 5 MHz readout. The measured read noise and FWHM for 2--5 MHz readout speeds are summarized in Table \ref{table:results}.

\clearpage
\subsection{MCRC ASIC performance}
\label{sect:asicperf}

The readout ASIC development and testing has been underway for several years at Stanford University, with the current MCRC v1.0 showing excellent performance \cite{Herrmannetal2020,Chattopadhyayetal2020,Oreletal2022,Herrmannetal2022,Chattopadhyayetal2022}, as summarized in Table \ref{tab:MCRCV1}. The integrated readout chip outperforms the established discrete electronics in all parameters, at a fraction of the power consumption.

\begin{table}[t]
\caption{Overview table of the MCRC V1.0 readout ASIC performance }\label{tab:MCRCV1}
\begin{center}       
\begin{tabular}{|l|l|} 
\hline\hline
Number of channels            & 8 channels / chip    \\\hline 
Input capacitance             & 1.5 pF \\\hline
Achievable pixel rate         & 5 Mpix/s per channel \\\hline
Voltage gain                  & 6.2 (low gain mode) \\
                              & 12.1 (high gain mode) \\\hline
Input dynamic range           & 320 mV (equivalent to ~30 keV) \\\hline
Input noise density           & 6.5 nV/$\sqrt{\mathrm{Hz}}$ \\\hline
Crosstalk                     & $\leq$ $-75$ dB in passband  \\\hline
Power consumption             & 35 mW per channel \\\hline
Radiation tolerance           & $\geq$ 25 krad  \\
\hline\hline
\end{tabular}
\end{center}
\end{table} 

\begin{figure}[t]
\begin{center}
\includegraphics[width=.45\linewidth]{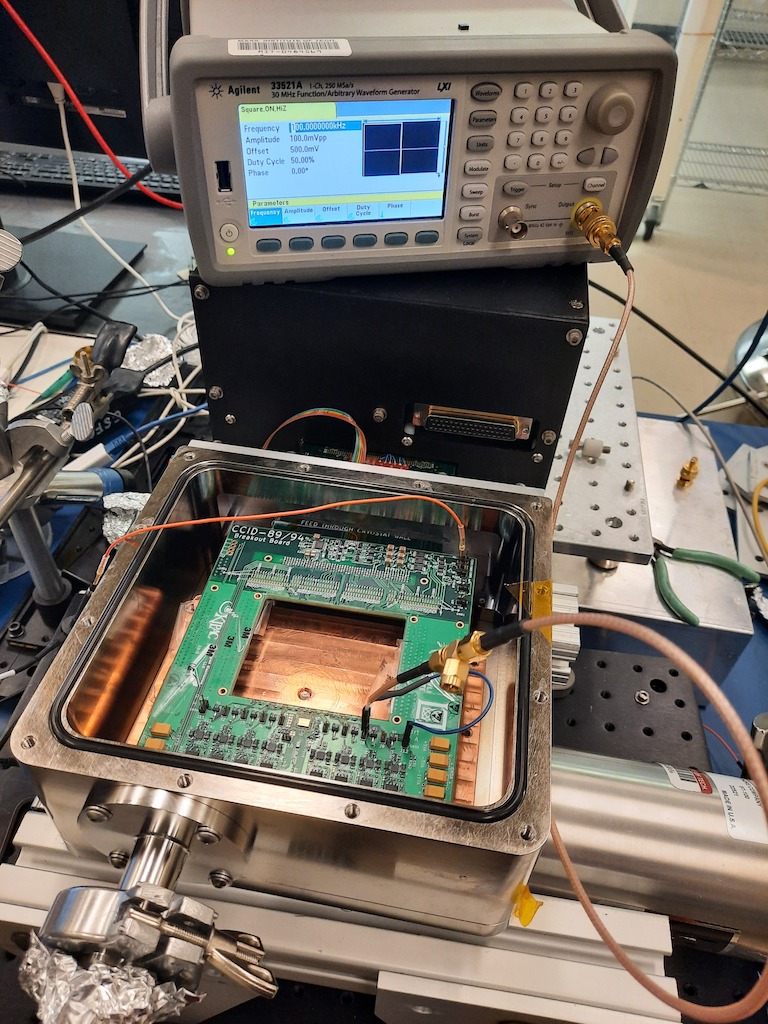}
\hspace*{.1in}
\includegraphics[width=.36\linewidth]{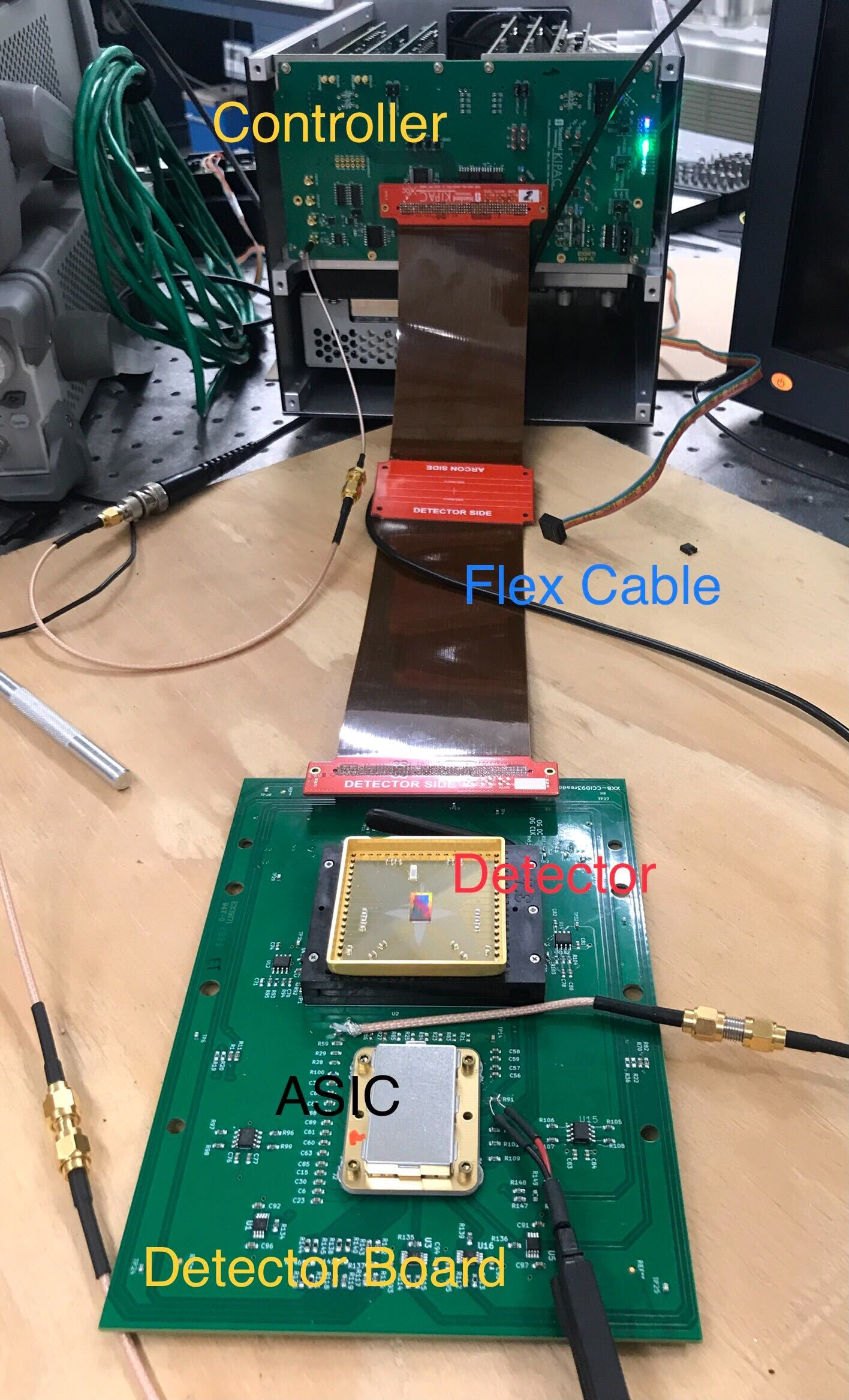}
\end{center}
\caption{(left) MCRC ASIC mounted on an MIT/LL CCID-89 laboratory package (without a CCD) for testing at MKI. This assembly supports high-speed operation of eight channels in parallel. (right) The MCRC ASIC and CCID-93 test setup at Stanford. This setup supports single channel operation, with excellent flexibility and debugging capability. It also prototypes a new high density flex lead, a precursor to the AXIS flex lead design.}
\label{fig:ccdasictest}
\end{figure}

Recently, the group has assembled the first combination of the MCRC with an MIT/LL prototype CCD package at MKI, and tested for functionality and performance across all eight ASIC readout channels. 
The anticipated speed, power consumption, and baseline noise were all achieved, with the MCRC clearly outperforming the previous discrete electronics.  In parallel, the group is building a second setup at Stanford University utilizing the smaller (and more economical) single-channel MIT/LL CCID-93 CCDs, to further characterize the combined ASIC+CCD performance and develop procedures for optimum operation. The setup will also be used to drive further development of the readout chip. Preliminary results at room temperature with the setup show near identical detector video waveforms for both the discrete and ASIC readout modes. The two ASIC enabled setups are shown in Figure \ref{fig:ccdasictest}.

We have recently tested a set of MCRC chips at the Brookhaven National Laboratory radiation test facility, exposing them to gamma rays of approximately 1 MeV (from a $^{60}$Co source) with a total ionizing dose of roughly 50 Mrad. All of the chips exhibited anticipated deviations of their operating points above 25 Mrad but no chip failed or showed any measurable degradation in the monitored response. The MCRC has thus been proven to be a successful, robust all-in-one readout solution.

We will soon be ready to pair a large, representative CCD (CCID-89) with the MCRC ASIC and verify the full X-ray performance.

\section{Summary}
\label{sect:summary}

The high throughput and excellent spatial resolution of AXIS will enable breakthrough science in the next decade. However, these exceptional performance characteristics also present technical challenges to design and fabricate a fully complementary X-ray camera. We exploit recent technology advances from MIT/LL that allow us to use well-established CCD technology in combination with a newly developed ASIC-based readout chip. These advances provide the speed required while also achieving the spectral performance necessary to realize the full potential of AXIS. The detector electronics combine advanced digital processing techniques with proven flight hardware and firmware design. The advanced detectors are housed in a high-heritage camera that builds on decades of flight experience and the associated lessons learned. Performance testing of prototype CCDs and ASICs produces excellent results, with low noise and good spectral response at the speeds required for AXIS. We have recently mated ASICs to our CCD test packages with excellent preliminary results, and technical demonstration of our advanced detector system continues on schedule to meet the needs of AXIS.

\acknowledgments 

We gratefully acknowledge support from NASA through the Strategic Astrophysics Technology (SAT) program, grants 80NSSC18K0138 and 80NSSC19K0401 to MIT, and from the Kavli Research Infrastructure Fund of the MIT Kavli Institute for Astrophysics and Space Research. Stanford team members acknowledge support from NASA through Astrophysics Research and Analysis (APRA) grants 80NSSC19K0499 and 80NSSC22K1921, and from the Kavli Institute for Particle Astrophysics and Cosmology.

\bibliography{edm} 
\bibliographystyle{spiebib} 

\end{document}